\documentclass{aa}

\usepackage{booktabs}
\usepackage{threeparttable}
\usepackage{graphicx}
\usepackage{multirow}
\usepackage{txfonts}
\usepackage{rotating}
\usepackage{hyperref}

\begin{document}
\title{Membership determination of 45 open clusters beyond 3 kpc}
\titlerunning{aa61340-26}

\author{
    Qian~Cui\inst{1},
    Zhihong~He\inst{1}
    \and
    Yangping~Luo\inst{1}
    \and Yan Jiang \inst{1}
}

\institute{
    School of Physics and Astronomy, China West Normal University, No.~1 Shida Road, Nanchong 637002, China\\
    \email{hezh@mail.ustc.edu.cn}
}

\date{Received 9 June 2026; accepted 30 August 2026}

\abstract
{Open clusters (OCs) are fundamental tracers for studying the structure and chemical evolution of the Galactic disc. Reliable identification of cluster members, particularly at the faint end of the main sequence, remains challenging owing to severe field-star contamination and the declining photometric completeness for distant, obscured systems.}
{We aim to identify and characterise distant OC candidates at heliocentric distances $d \gtrsim 3$~kpc using \textit{Gaia} DR3 astrometry, with emphasis on improving the recovery of faint candidate members in highly contaminated stellar fields.}
{We apply a probabilistic membership method based on \textit{Gaia} DR3 proper motions and parallaxes. Candidate members are selected through combined astrometric criteria and refined via spatial filtering using radial density profiles, with photometric data serving as a consistency check. For clusters affected by significant extinction, near-infrared photometry from 2MASS is incorporated to compensate for the limitations of optical data.}
{We analyse a sample of 45 targets, comprising 30 previously reported OCs and 15 newly identified candidates, all located at $d \gtrsim 3$~kpc. The method yields more compact astrometric distributions and improves the recovery of faint candidate members in several systems. However, residual field contamination remains non-negligible for the most obscured clusters projected against the Galactic plane.} 
{\textit{Gaia} DR3 astrometry provides an effective basis for the exploratory identification and first-order characterisation of distant OC candidates. Dedicated near-infrared follow-up observations will be essential for robust membership assessment in high extincted regions.}

\maketitle

\section{Introduction}\label{sec:intro}

	Open clusters (OCs) are well-established tracers of the Galactic disc \citep{Bland-Hawthorn10,Magrini17,Spina21}. Their member stars share a common origin and therefore have similar ages, distances, and initial chemical compositions \citep{Lada03}. Given their wide distribution across the Milky Way and broad age range, OCs place important constraints on the structure, kinematics, and chemical evolution of the Galactic disc \citep{JA1982,Friel1995,Soubiran2018,CG20,he23c}. The \textit{Gaia} mission has substantially advanced this field by providing homogeneous astrometric and photometric data of unprecedented precision \citep{GaiaDR2,Brown21,GaiaDR3}. These data have improved both the discovery of new clusters and the characterisation of known systems, leading to significant revisions of the Galactic cluster census and more homogeneous determinations of membership and fundamental parameters \citep[e.g.][]{CG18,Castro20,Castro22,LP19,Bossini2019,Ferreira20,Ferreira21,Dias21,Dias25,HR23,Cavallo24,Sim19,he21a,he22b,He23b}. 
	
	Despite this progress, the identification of distant OCs remains challenging. At large heliocentric distances, clusters subtend smaller angular sizes on the sky and their members become increasingly mixed with surrounding field stars, reducing the contrast of the overdensity in both positional and astrometric space \citep{CG20,HR23}. Many distant clusters are also located close to the Galactic plane, where strong and inhomogeneous interstellar extinction reduces source brightness and further complicates both astrometric and photometric analyses \citep{Carraro07,Castro21}. These effects are particularly severe for faint members, for which the astrometric quality of \textit{Gaia} data degrades with source magnitude and colour, and for which systematic effects such as the parallax bias become non-negligible \citep{Brown21,Lindegren2021a}. Distant OCs therefore require careful treatment when used as tracers of the large-scale Galactic structure \citep{Perren22}.

    The membership determination is a critical step in the OC study. Early work relied mainly on kinematic information, most notably the classical Vasilevskis--Sanders (VS) method, in which cluster members and field stars are modelled as separate bivariate Gaussian distributions in proper-motion space \citep{Vasilevskis1958,Sander1971,Zhao1990}. Such parametric approaches become less effective when the cluster-to-field contrast is low or when the field-star distribution departs significantly from a Gaussian profile \citep{cabrea_alfaro1990,Balaguer2004}. To relax these assumptions, a variety of non-parametric approaches were developed. \citet{cabrea_alfaro1990} constructed empirical probability density functions directly from kinematic observables, while \citet{Balaguer2004} applied kernel-density estimation (KDE) in proper-motion space. \citet{Javakhishvili2006} introduced minimum relative-velocity grouping to enhance the contrast between cluster members and field stars. The UPMASK algorithm \citep{krone2014} further advanced this approach by combining random resampling with minimum-spanning-tree voting, producing membership labels without user-defined thresholds and showing robustness against non-Gaussian or multimodal field distributions.

    Before the \textit{Gaia} era, two widely used reference catalogues were the compilation of \citet[][hereafter Dias02]{Dias02} and the Milky Way Star Clusters catalogue of \citet[][hereafter K13]{Kharchenko13}. Dias02 compiled positional, kinematic, and astrophysical information for known OCs from the literature and was periodically updated as new measurements became available. K13 provided a comparatively homogeneous redetermination of cluster membership and fundamental parameters, based mainly on 2MASS photometry and PPMXL astrometry. These catalogues served as important reference samples for OC studies before the availability of precise \textit{Gaia} astrometry.

    The \textit{Gaia} mission has enabled the application of higher-dimensional clustering algorithms and machine-learning techniques to OC membership determination. Recent studies have employed methods such as spectral clustering, $k$-nearest-neighbour algorithms, Gaussian mixture models, and neural networks in astrometric or combined astrometric--photometric spaces, leading to substantial improvements in member recovery \citep{Gao2018,Agarwal2021,deb2022,VMG23}. Despite these advances, the identification of distant OCs and their faint members remains difficult. In the low-contrast regime typical of sparse and distant systems, current methods may still suffer from incomplete member recovery and reduced robustness, particularly in regions of strong extinction and heavy field contamination.

   Recent \textit{Gaia}-based catalogues have greatly expanded the census of Galactic OCs. \citet[][hereafter CG20]{CG20} presented one of the largest homogeneous cluster samples derived from \textit{Gaia} data, which has since served as a standard reference in the field. However, a fraction of their distant objects have only a small number of identified members, as indicated by quality flags such as \texttt{tooRed} or \texttt{notEnoughStars}. These systems are difficult to validate both structurally and photometrically, and some lack independent confirmation in other recent compilations \citep[e.g.][]{Dias21,HR23}. \citet[][hereafter Dias21]{Dias21} compiled Gaia DR2-based membership determinations and derived a homogeneous set of astrometric and fundamental parameters for 1743 OCs, while \citet[][hereafter HR23]{HR23} performed a blind all-sky search using Gaia DR3 and provided new membership lists and cluster parameters. Both catalogues contain subsets of the 30 distant OCs considered here and therefore provide useful external references for assessing our membership determinations and derived cluster parameters. These clusters remain tentative and require further verification, motivating the analyses presented in this study.

   In this work, we revisit 30 distant OCs from the CG20 catalogue using \textit{Gaia} DR3 astrometry, reassessing both stellar membership and fundamental parameters. Applying this approach to fields from our previous OC study \citep{He23b} leads to the identification of 15 new OC candidates, predominantly located at heliocentric distances greater than 4 kpc. These results help to extend the census of star clusters in the outer Galactic disc.

   This paper is organised as follows. Section~\ref{sec:data} describes the \textit{Gaia} DR3 data and the distant OC sample used in this work. Section~\ref{sec:analysis} presents the membership determination procedure and the structural analysis, including the radial density profile fitting used to estimate cluster radii. Section~\ref{sec:results} reports the results for the distant cluster sample and the newly identified candidates. Section~\ref{sec:conclusions} summarises the main findings and outlines directions for future work.

\section{Data}\label{sec:data}

\subsection{\textit{Gaia} DR3}\label{Gaia}

\textit{Gaia} DR3 \citep[][]{GaiaDR3} provides astrometric and photometric measurements for more than one billion sources, together with radial velocities for approximately 33.8 million stars. The astrometric precision degrades towards fainter magnitudes. For bright sources ($G < 15$~mag), typical uncertainties are $0.01$--$0.02$~mas in position, $0.02$--$0.03$~mas in parallax, and $0.02$--$0.03$~mas~yr$^{-1}$ in proper motion. At $G \approx 21$~mag, these increase to approximately 1.0~mas, 1.3~mas, and 1.4~mas~yr$^{-1}$, respectively. 

In this work, \textit{Gaia}~DR3 parallaxes were corrected for the global zero-point offset following \citet{Lindegren2021a}, and the corrected values form the basis of all subsequent astrometric analyses. We also account for the significant large-scale angular covariance present in \textit{Gaia} parallaxes and proper motions, which causes the uncertainties of mean cluster quantities to deviate from the simple $1/\sqrt{N}$ scaling. Covariance terms are incorporated following \citet{Lindegren2021b}; for typical DR3 fields, their amplitudes are of order $700\,\mu\mathrm{as}^2$ for parallaxes and $550\,\mu\mathrm{as}^2\,\mathrm{yr}^{-2}$ for proper motions. These terms are included when propagating uncertainties in the astrometric metrics used for membership determination.

\subsection{Distant OCs catalogue}\label{sec:catalogue}

Our sample is from the CG20 catalogue and consists of 30 distant OCs with heliocentric distances $d > 3$~kpc ($\varpi \lesssim 0.33$~mas). These clusters were selected based on flags indicating severe extinction (\texttt{tooRed}) or an insufficient number of identified members (\texttt{notEnoughStars}). For context, we cross-matched our sample against several independent catalogues: the \textit{Gaia}-based compilations of HR23 and Dias21, as well as the pre-\textit{Gaia} catalogues of Dias02 and K13. Cross-matching was performed using both sky coordinates and cluster identifiers. Of the 30 clusters, 26 have counterparts in Dias02, 24 in K13, 17 in HR23, and 5 in Dias21. These published parameters (Table~\ref{tab:age_dist_30ocs}) are not homogenised, reflecting differences in membership criteria and isochrone fitting, which further justifies the uniform re-analysis presented below.

\begin{table*}[htbp]
\centering
\footnotesize
\caption{Comparison of cluster parameters from different catalogues.}
\label{tab:age_dist_30ocs}

\begin{threeparttable}
\resizebox{\textwidth}{!}{
\begin{tabular}{lcccccccccccccccccc}
\toprule
\multirow{2}{*}{Cluster} & \multirow{2}{*}{$\alpha$ [deg]} & \multirow{2}{*}{$\delta$ [deg]} &
\multicolumn{3}{c}{CG20} &
\multicolumn{3}{c}{Dias02} &
\multicolumn{3}{c}{K13} &
\multicolumn{3}{c}{Dias21} &
\multicolumn{3}{c}{HR23} \\
\cmidrule(lr){4-6}\cmidrule(lr){7-9}\cmidrule(lr){10-12}\cmidrule(lr){13-15}\cmidrule(lr){16-18}
 & & &
$N$ & $\log (Age/yr)$ &  $Dist. $  &
$N$ & $\log (Age/yr)$ &  $Dist. $  &
$N$ & $\log (Age/yr)$ &  $Dist. $  &
$N$ & $\log (Age/yr)$ &  $Dist. $  &
$N$ & $\log (Age/yr)$ &  $Dist. $  \\
\midrule

        Arp\_Madore\_2 & 114.683 & -33.845 & 29 & 9.48 & 11751 & 19 & 9.34 & 13341 & - & - & - & - & - & - & - & - & - \\
        Berkeley\_43 & 288.872 & 11.265 & 197 & - & - & 6 & 8.50 & 1030 & 293 & 8.79 & 1355 & 204 & 8.31 & 2459 & 472 & 7.88 & 3433 \\ 
        Berkeley\_51 & 302.972 & 34.402 & 89 & - & - & 101 & 9.05 & 1300 & 326 & 8.26 & 3292 & - & - & - & 159 & 7.69 & 4969 \\ 
        Berkeley\_52 & 303.621 & 28.947 & 63 & - & - & 5 & 9.30 & 4900 & 171 & 9.00 & 4947 & - & - & - & 159 & 9.38 & 5274 \\ 
        BH\_222 & 259.694 & -38.289 & 104 & - & - & 27 & 7.78 & 6000 & - & - & - & - & - & - & 63 & 7.16 & 2147 \\ 
        DC\_5 & 159.996 & -59.192 & 125 & - & - & 37 & 6.05 & 4430 & 88 & 6.05 & 4430 & - & - & - & 135 & 7.27 & 4422 \\ 
        ESO\_211\_09 & 139.186 & -50.285 & 27 & - & - & 21 & 9.05 & 5823 & - & - & - & - & - & - & - & - & - \\ 
        FSR\_0975 & 100.010 & 13.310 & 7 & - & - & 77 & 9.10 & 4626 & 26 & 9.10 & 4626 & - & - & - & 86 & 8.75 & 4364 \\ 
        FSR\_1025 & 102.620 & 6.612 & 10 & 8.75 & 6136 & 8 & 8.60 & 2095 & 43 & 8.60 & 2095 & - & - & - & - & - & - \\ 
        FSR\_1171 & 105.058 & -10.320 & 15 & - & - & 44 & 9.50 & 4110 & 44 & 9.20 & 4933 & - & - & - & 91 & 8.37 & 3468 \\ 
        Ivanov\_8 & 124.794 & -35.658 & 14 & - & - & 55 & 9.20 & 3888 & 55 & 7.80 & 1194 & - & - & - & 24 & 8.77 & 4507 \\ 
        Kronberger\_79 & 293.478 & 18.522 & 59 & - & - & 56 & 8.35 & 2700 & 56 & 8.39 & 3369 & 63 & 6.76 & 3296 & 53 & 7.04 & 4790 \\ 
        Kronberger\_85 & 119.590 & -34.770 & 29 & - & - & 39 & 8.80 & 8500 & 39 & 8.50 & 10380 & - & - & - & - & - & - \\ 
        Mayer\_3 & 112.523 & -18.553 & 23 & - & - & 38 & 7.10 & 2494 & - & - & - & - & - & - & 19 & 6.97 & 3777 \\ 
        NGC\_1624 & 70.162 & 50.458 & 10 & - & - & 149 & 6.60 & 6000 & 81 & 6.65 & 5466 & - & - & - & 24 & 7.32 & 4261 \\ 
        NGC\_3603 & 168.795 & -61.259 & 272 & - & - & 12 & 6.00 & 6900 & 29 & 6.00 & 7158 & - & - & - & 215 & 6.85 & 6775 \\ 
        Schuster\_1 & 151.159 & -55.857 & 17 & - & - & 9 & 8.35 & 5212 & 37 & 6.80 & 1863 & - & - & - & - & - & - \\ 
        Shorlin\_1 & 166.443 & -61.231 & 2 & - & - & - & - & - & 38 & 6.50 & 5594 & - & - & - & - & - & - \\ 
        Westerlund\_1 & 251.760 & -45.852 & 308 & - & - & 13 & 6.70 & 5500 & 13 & 6.90 & 7686 & - & - & - & - & - & - \\ 
        Westerlund\_2 & 156.010 & -57.758 & 168 & - & - & 51 & 6.30 & 2850 & 51 & 6.60 & 4469 & - & - & - & 101 & 7.88 & 4533 \\
        Alessi\_59 & 103.078 & 2.196 & 10 & - & - & 33 & 7.60 & 3500 & 95 & 7.00 & 3511 & 11 & 8.76 & 4387 & 123 & 8.14 & 3820 \\ 
        ESO\_313\_03 & 127.907 & -41.784 & 3 & - & - & 13 & 8.85 & 3048 & - & - & - & - & - & - & - & - & - \\ 
        FSR\_0524 & 14.340 & 62.125 & 8 & - & - & - & - & - & 101 & 9.19 & 1841 & 18 & 8.44 & 3594 & 64 & 8.34 & 3555 \\ 
        FSR\_1580 & 171.917 & -58.471 & 7 & - & - & - & - & - & 38 & 8.90 & 2511 & - & - & - & - & - & - \\ 
        Juchert\_1 & 290.629 & 12.667 & 53 & - & - & 10 & 9.10 & 2623 & 142 & 8.68 & 2509 & 60 & 8.59 & 3586 & 108 & 7.86 & 3263 \\ 
        Kronberger\_54 & 300.778 & 31.968 & 10 & - & - & 105 & 8.40 & 1715 & 105 & 8.83 & 2900 & - & - & - & 72 & 7.03 & 4025 \\ 
        Patchick\_94 & 247.397 & -47.310 & 25 & - & - & 17 & 9.07 & 4126 & 60 & 6.55 & 3178 & - & - & - & - & - & - \\ 
        Ruprecht\_32 & 116.300 & -25.539 & 13 & - & - & 88 & 7.08 & 5346 & 57 & 6.70 & 4417 & - & - & - & - & - & - \\ 
        SAI\_72 & 103.973 & 0.232 & 6 & - & - & 54 & 8.50 & 3150 & 57 & 8.49 & 3239 & - & - & - & - & - & - \\ 
        UBC\_431 & 75.222 & 41.232 & 72 & - & - & - & - & - & - & - & - & - & - & - & - & - & - \\  

\bottomrule
\end{tabular}
} 

\begin{tablenotes}[flushleft]
\footnotesize
\item \parbox{\textwidth}{\textit{Note.} Columns are as follows: 1) cluster name; 2) right ascension ($\alpha$) in degrees; 3) declination ($\delta$) in degrees. For each catalogue, $N$ denotes the number of stars used to derive the cluster parameters, $\log (Age/yr)$ is the logarithm of the cluster age in years, and $Dist.$ is the heliocentric distance in parsecs. The distance values are adopted directly from the corresponding catalogues and are not derived from the inverse Gaia parallaxes in this work.}
\end{tablenotes}

\end{threeparttable}
\end{table*}

\begin{figure*}
    \centering
    \includegraphics[width=0.85\linewidth]{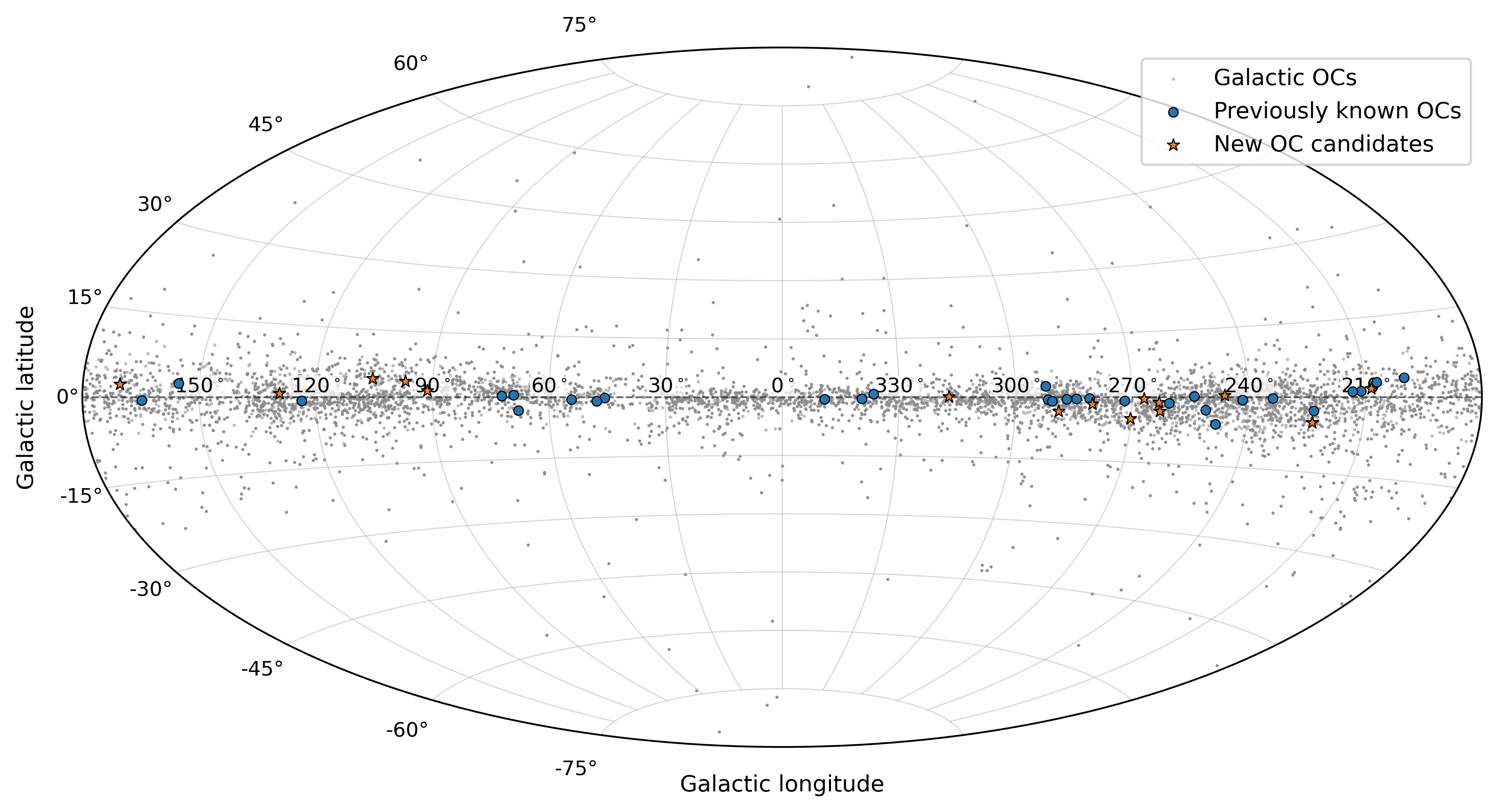}
    \caption{Galactic distribution of the 45 OCs. Grey points show the catalogued Galactic star clusters from \citet{He23b} and \citet{HR23}, while blue circles and orange stars denote the 30 previously known clusters and 15 newly identified OC candidates, respectively. The dashed line marks the Galactic plane.}
    \label{fig:lb_45oc}
\end{figure*}

The Galactic distribution of the 45 OCs is shown in Figure~\ref{fig:lb_45oc}. The sample includes 30 previously known OCs and 15 newly identified candidates, shown together with the catalogued Galactic OC population for comparison. Most of the targets are concentrated towards the Galactic plane but cover a broad range of Galactic longitudes.

\section{Membership identification} \label{sec:analysis}

For each cluster, \textit{Gaia} DR3 sources were retrieved within a cone of radius 45~arcmin centred on the catalogue position \footnote{The adopted radius was motivated by the structural properties of distant OCs reported by \citet{Cui25}, whose tidal radii typically span $\sim10$--$25$~pc, with a median of $\sim17.5$~pc. At a distance of 4~kpc, three times the median tidal radius corresponds to approximately 45~arcmin. Adopting $3r_{\rm t}\sim75$~pc, based on the upper end of the reported tidal-radius range, would produce an unnecessarily large extraction region for most clusters and introduce additional unrelated field stars.}. The 45~arcmin cone was adopted as a homogeneous initial extraction region intended to include both the cluster population and a representative surrounding field-star sample. No additional cuts on astrometric quality indicators, such as RUWE or proper-motion uncertainties, were applied at this stage to avoid discarding faint candidate members.

\subsection{Membership probability} \label{sec:prob}

For distant OCs, a large fraction of the member stars lie at the faint end of the \textit{Gaia} sample, where astrometric uncertainties are larger. Combined with the low cluster-to-field contrast typical of sparse and distant systems, this makes the recovery of cluster signals less straightforward. High-dimensional clustering methods can in such cases become sensitive to measurement noise, parameter choices, and local density fluctuations \citep{Balaguer2004,krone2014,HR23}. We therefore adopt a probabilistic approach based on the VS model, which provides a simple and homogeneous framework well suited to the analysis of distant and sparse clusters. The same method was applied in our previous work \citep{Cui25}, to which we refer the reader for a detailed description. In brief, membership probabilities are computed from the joint distribution of proper motions ($\mu_{\alpha}^{*}$, $\mu_{\delta}$) and parallax ($\varpi$), using the zero-point-corrected astrometry and covariance-adjusted uncertainties described in Section~\ref{Gaia}. Stars with $P \geq 0.2$ are retained as a preliminary candidate sample for the structural analysis. This permissive threshold is adopted only to define the cluster radial density profile and estimate the cluster radius, while the final membership selection is performed as described in Section~\ref{sec:selection}.

\subsubsection{Proper motion}
The cluster and field populations are described by two-dimensional Gaussian distributions in the proper-motion plane. Both components are allowed to have different dispersions along the two proper-motion axes, while the cluster distribution is generally more compact than the field distribution. Their frequency distributions are given by

\begin{align}
\phi_{\rm c\mu}(i) &=
\frac{N_{\rm c\mu}}
{2\pi\sqrt{1-\gamma_{{\rm c},i}^{2}}
\sqrt{
(\sigma_{\alpha*,{\rm c}}^{2}+\epsilon_{\alpha*,i}^{2})
(\sigma_{\delta,{\rm c}}^{2}+\epsilon_{\delta,i}^{2})
}}
\nonumber\\
&\quad \times
\exp\Bigg\{
-\frac{1}{2(1-\gamma_{{\rm c},i}^{2})}
\Bigg[
\frac{(\mu_{\alpha*,i}-\mu_{\alpha*,{\rm c}})^{2}}
{\sigma_{\alpha*,{\rm c}}^{2}+\epsilon_{\alpha*,i}^{2}}
+
\frac{(\mu_{\delta,i}-\mu_{\delta,{\rm c}})^{2}}
{\sigma_{\delta,{\rm c}}^{2}+\epsilon_{\delta,i}^{2}}
\nonumber\\
&\quad -
\frac{
2\gamma_{{\rm c},i}
(\mu_{\alpha*,i}-\mu_{\alpha*,{\rm c}})
(\mu_{\delta,i}-\mu_{\delta,{\rm c}})
}{
\sqrt{
(\sigma_{\alpha*,{\rm c}}^{2}+\epsilon_{\alpha*,i}^{2})
(\sigma_{\delta,{\rm c}}^{2}+\epsilon_{\delta,i}^{2})
}}
\Bigg]
\Bigg\}.
\end{align}

and

\begin{align}
\phi_{\rm f\mu}(i) &=
\frac{N_{\rm f\mu}}
{2\pi\sqrt{1-\gamma_{{\rm f},i}^{2}}
\sqrt{
(\sigma_{\alpha*,{\rm f}}^{2}+\epsilon_{\alpha*,i}^{2})
(\sigma_{\delta,{\rm f}}^{2}+\epsilon_{\delta,i}^{2})
}}
\nonumber\\
&\quad \times
\exp\Bigg\{
-\frac{1}{2(1-\gamma_{{\rm f},i}^{2})}
\Bigg[
\frac{(\mu_{\alpha*,i}-\mu_{\alpha*,{\rm f}})^{2}}
{\sigma_{\alpha*,{\rm f}}^{2}+\epsilon_{\alpha*,i}^{2}}
+
\frac{(\mu_{\delta,i}-\mu_{\delta,{\rm f}})^{2}}
{\sigma_{\delta,{\rm f}}^{2}+\epsilon_{\delta,i}^{2}}
\nonumber\\
&\quad -
\frac{
2\gamma_{{\rm f},i}
(\mu_{\alpha*,i}-\mu_{\alpha*,{\rm f}})
(\mu_{\delta,i}-\mu_{\delta,{\rm f}})
}{
\sqrt{
(\sigma_{\alpha*,{\rm f}}^{2}+\epsilon_{\alpha*,i}^{2})
(\sigma_{\delta,{\rm f}}^{2}+\epsilon_{\delta,i}^{2})
}}
\Bigg]
\Bigg\}.
\end{align}

The proper-motion membership probability of the $i$-th star is then calculated as

\begin{equation}
P_{\mu}(i)=
\frac{\phi_{\rm c\mu}(i)}
{\phi_{\rm c\mu}(i)+\phi_{\rm f\mu}(i)}.
\end{equation}

Here, $\mu_{\alpha*,i}$ and $\mu_{\delta,i}$ denote the proper-motion components of the $i$-th star in right ascension and declination, respectively, where $\mu_{\alpha*}=\mu_{\alpha}\cos\delta$. Their corresponding measurement errors are denoted by $\epsilon_{\alpha*,i}$ and $\epsilon_{\delta,i}$. The quantities $(\mu_{\alpha*,{\rm c}},\mu_{\delta,{\rm c}})$ and $(\mu_{\alpha*,{\rm f}},\mu_{\delta,{\rm f}})$ represent the median proper-motion components of the cluster and field populations, respectively. $\sigma_{\alpha*,{\rm c}}$ and $\sigma_{\delta,{\rm c}}$ denote the proper-motion dispersions of the cluster population along the two proper-motion axes, while $\sigma_{\alpha*,{\rm f}}$ and $\sigma_{\delta,{\rm f}}$ denote the corresponding dispersions of the field population. The normalisation coefficients satisfy $N_{\rm c\mu}+N_{\rm f\mu}=1$.

The quantity $\rho_i= {pmra\_pmdec\_corr}_i$ is the correlation coefficient between the \textit{Gaia} DR3 proper-motion measurement errors of the $i$-th star. The corresponding measurement-error covariance is $\rho_i\epsilon_{\alpha*,i}\epsilon_{\delta,i}$. The effective correlation terms entering the cluster and field distributions are therefore

\begin{align}
\gamma_{{\rm c},i}
&=
\frac{
\rho_i\epsilon_{\alpha*,i}\epsilon_{\delta,i}
}{
\sqrt{
(\sigma_{\alpha*,{\rm c}}^{2}+\epsilon_{\alpha*,i}^{2})
(\sigma_{\delta,{\rm c}}^{2}+\epsilon_{\delta,i}^{2})
}
},
\\
\gamma_{{\rm f},i}
&=
\frac{
\rho_i\epsilon_{\alpha*,i}\epsilon_{\delta,i}
}{
\sqrt{
(\sigma_{\alpha*,{\rm f}}^{2}+\epsilon_{\alpha*,i}^{2})
(\sigma_{\delta,{\rm f}}^{2}+\epsilon_{\delta,i}^{2})
}
}.
\end{align}

Thus, $\gamma_{{\rm c},i}$ and $\gamma_{{\rm f},i}$ represent the effective correlation terms for the cluster and field distributions, respectively. In both cases, the off-diagonal covariance arises solely from the \textit{Gaia} proper-motion measurement covariance; no additional intrinsic cross-covariance is assumed.

\subsubsection{Parallax}

Following the same approach as for proper motions, the parallax distributions of the cluster and field populations are modelled as two Gaussian components. Their frequency distributions are given by

\begin{align}
\phi_{\rm c\varpi}(i) &=
\frac{N_{\rm c\varpi}}
{2\pi\sqrt{\sigma_{\rm c\varpi}^{2}+\epsilon_{\varpi,i}^{2}}}
\times
\exp\left\{
-\frac{1}{2}
\left[
\frac{(\varpi_i-\varpi_{\rm c})^{2}}
{\sigma_{\rm c\varpi}^{2}+\epsilon_{\varpi,i}^{2}}
\right]
\right\},
\end{align}

and

\begin{align}
\phi_{\rm f\varpi}(i) &=
\frac{N_{\rm f\varpi}}
{2\pi\sqrt{\sigma_{\rm f\varpi}^{2}+\epsilon_{\varpi,i}^{2}}}
\times
\exp\left\{
-\frac{1}{2}
\left[
\frac{(\varpi_i-\varpi_{\rm f})^{2}}
{\sigma_{\rm f\varpi}^{2}+\epsilon_{\varpi,i}^{2}}
\right]
\right\}.
\end{align}

The parallax-based membership probability of the $i$-th star is then

\begin{equation}
P_{\varpi}(i)=
\frac{\phi_{\rm c\varpi}(i)}
{\phi_{\rm c\varpi}(i)+\phi_{\rm f\varpi}(i)}.
\end{equation}

Here, $\varpi_i$ and $\epsilon_{\varpi,i}$ denote the parallax and its measurement errors for the $i$-th star, respectively. The quantities $\varpi_{\rm c}$ and $\varpi_{\rm f}$ represent the central parallaxes of the cluster and field populations, while $\sigma_{\rm c\varpi}$ and $\sigma_{\rm f\varpi}$ denote their corresponding dispersions. The normalisation coefficients satisfy $N_{\rm c\varpi}+N_{\rm f\varpi}=1$. The resulting probability $P_{\varpi}(i)$ represents the membership probability inferred from the parallax distribution alone.

\subsection{Radial density profile}\label{sec:rdp}

Following \citet{Cui25}, we construct the radial density profile (RDP) of each cluster in a uniform way. For each candidate member, the projected distance $r$ from the cluster centre is computed. The surrounding field is then divided into concentric annuli with a fixed radial step of 0.25~pc, and the surface stellar density in each annulus is estimated as $\rho = N/S$, where $N$ is the number of stars and $S$ is the corresponding area. 

From the resulting RDP, the core radius $r_{\rm c}$ is defined as the radius at which the surface density drops to half the central value $\rho_0$. The background density $\rho_{\rm bg}$ is estimated as the mean surface density in the outer region of the field, spanning 30\%--80\% of the maximum radial extent. This interval excludes the inner region, where cluster members still contribute significantly to the surface density, and the outermost bins, which may be affected by edge effects or spatial inhomogeneities. The dispersion of the background density is denoted $\sigma_{\rm bg}$.

We further define the cluster radius $r_{\rm clu}$ as the radius at which the surface density profile drops to 
\begin{equation} 
\rho(r_{\rm clu}) = \rho_{\rm bg} + 3\sigma_{\rm bg}, 
\end{equation} 
corresponding to a $3\sigma$ excess above the local background. We also define the density contrast parameter $\kappa = \rho_0 / \rho_{\rm bg}$, which quantifies the overdensity of the cluster centre relative to the surrounding field. Based on this parameter, we divide the sample into two groups: clusters with $\kappa > 3$ are classified as Type~I, indicating a statistically significant stellar overdensity, while those with $\kappa \leq 3$ are assigned to Type~II.

\subsection{Candidate selection}\label{sec:selection}
Cluster member selection follows a sequential procedure combining astrometric filtering, radial density analysis, and photometric validation.

First, proper-motion and parallax membership probabilities, $P_{\mu}$ and $P_{\varpi}$, are computed for all stars in the field. Stars with $P_{\mu} \geq 0.5$ and $P_{\varpi} \geq 0.5$ are retained as the initial candidate sample. This threshold represents a practical compromise between completeness and field-star contamination: a higher value would reduce contamination but risk removing faint members with noisy astrometry, while a lower value would retain more candidates at the cost of increased contamination in the subsequent analysis. The threshold should therefore be regarded as defining an initial candidate list for further refinement, rather than as a strict membership boundary. 

From these astrometric candidates, the RDP is constructed as described in Section~\ref{sec:rdp}, and only stars within $r \leq r_{\rm clu}$ are carried forward. This spatial cut ensures that the candidate sample is consistent with a genuine stellar overdensity rather than a chance projection of field stars. 

The colour--magnitude diagram (CMD) of the spatially selected candidates is then inspected visually to remove obvious photometric outliers that are inconsistent with the cluster sequence. This step eliminates residual field contaminants that passed the astrometric and spatial filters. The resulting member list forms the basis for the subsequent determination of cluster parameters. Fundamental parameters---age, extinction, and distance modulus---are derived by fitting PARSEC~1.2S isochrones \citep{Bressan2012} to the observed CMD of the final member sample. The extinction law adopted in the fitting is described in Appendix~\ref{app:extin_law}.

\section{Results and discussion}\label{sec:results} 

The membership identification strategy adopted in this work is designed to facilitate the identification of cluster members in distant and highly contaminated stellar fields. The combined use of astrometric, spatial, and photometric information can contribute to mitigate field-star contamination and may improve the recovery of faint candidate members. This is particularly relevant for distant OCs, whose lower main sequences are often difficult to trace owing to strong extinction, observational incompleteness, and severe background contamination. The resulting member catalogues generally exhibit coherent spatial and photometric distributions, providing a basis for the analysis presented in the following sections.

\subsection{Properties of the 30 known OCs} 
After re-determining the membership probabilities for the 30 OCs listed in CG20, we examine their structural and photometric properties to characterise the reliability of each system (Table~\ref{tab:this_work}). The RDP and CMD provide complementary diagnostics of the spatial concentration and evolutionary coherence of the candidate members. 

\begin{figure*}
    \centering
    \includegraphics[width=0.9\linewidth]{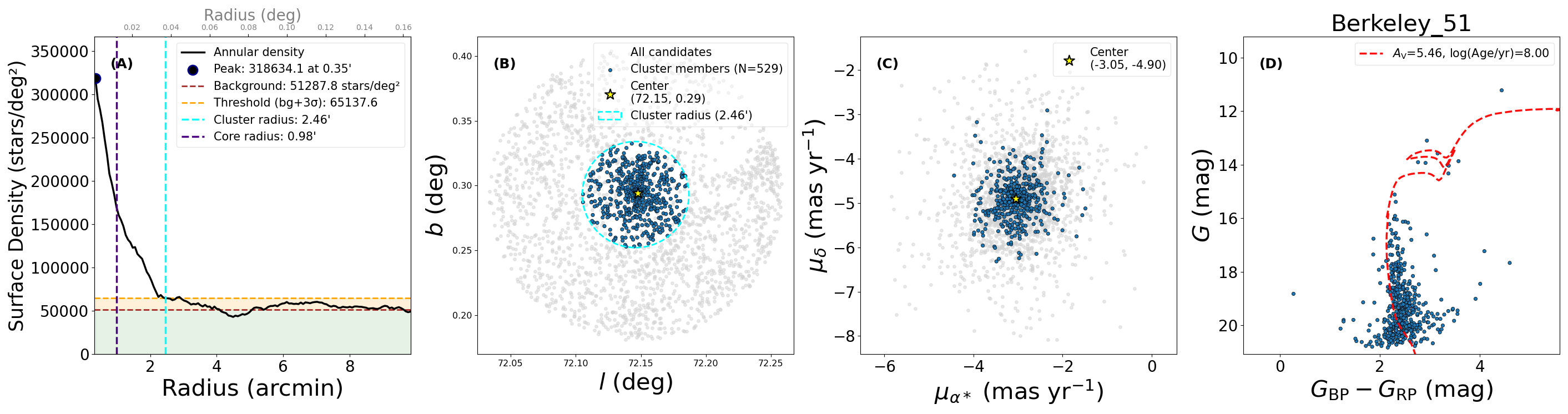}\\
    \includegraphics[width=0.9\linewidth]{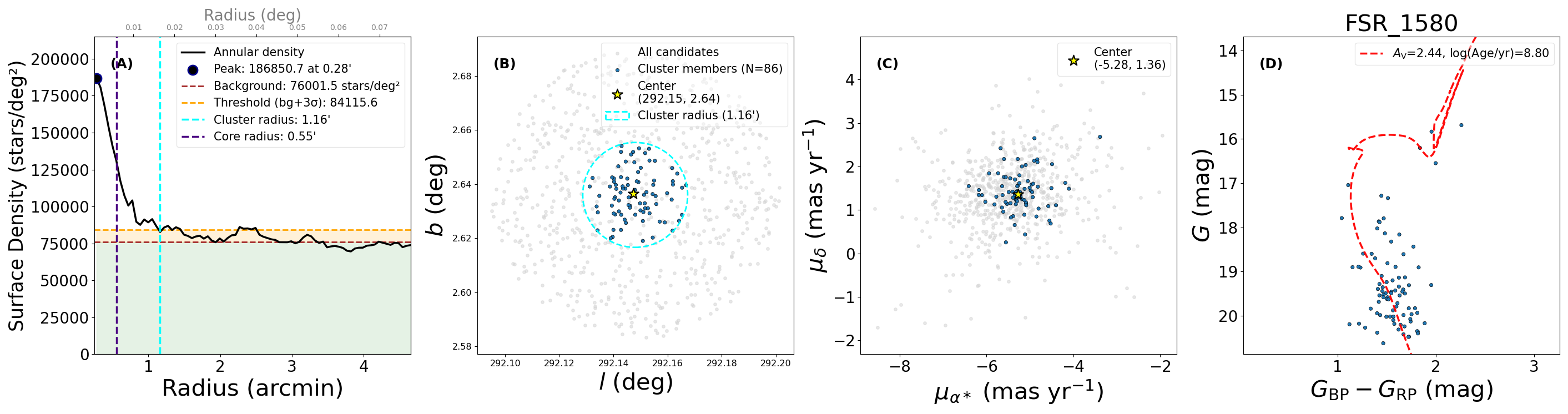}
    \caption{Structural and photometric analysis of Type I (top panel) and Type II (bottom panel) OCs. 
    Panel~(A) Radial density profile of the cluster region. The cluster radius ($r_{\rm clu}$) and core radius ($r_{\rm c}$) are indicated by cyan and indigo vertical lines, respectively. The brown and orange horizontal lines denote the background density and the density level $\rho(r_{\rm clu})$, respectively.
    Panel~(B) Spatial distribution of stars in Galactic coordinates $(l,\,b)$. Grey dots represent all candidate stars, while blue dots mark the selected cluster members. The cyan circle indicates the adopted cluster radius, and the yellow star marks the cluster centre.
    Panel~(C) Proper-motion diagram in the $(\mu_{\alpha*},\,\mu_{\delta})$ plane. 
    Grey points show field candidates and blue points indicate cluster members, with the yellow star denoting the mean proper motion of the cluster.
    Panel~(D) CMD of the selected cluster members. The red dashed line shows the adopted best-fitting isochrone, with the extinction $A_V$ and $\log(\mathrm{Age/yr})$ indicated in the panel. Specifically, Berkeley\_51, located towards the northern Galactic warp, exhibits a high concentration of member stars. In contrast, FSR\_1580 displays poor concentration. }
    \label{fig:reidenti_oc}
\end{figure*}

Of the 30 clusters, 21 show a clear central overdensity relative to the surrounding field ($\kappa > 3$, Type~I). For these systems, the RDPs display a smooth radial decline, and the CMDs show a continuous main sequence consistent with a single stellar population. The agreement between the spatial concentration and the photometric morphology supports the interpretation that these systems are genuine physical clusters rather than chance alignments of field stars. The remaining 9 clusters (Type~II, $\kappa \leq 3$) show weaker overdensities and less well-defined CMD sequences. An example is shown for Berkeley~51 (Type I) and FSR~1580 (Type II) in Figure~\ref{fig:reidenti_oc}; the corresponding figures for all clusters are available via Zenodo at \href{https://doi.org/10.5281/zenodo.22240669}{https://doi.org/10.5281/zenodo.22240669}.  

\subsection{Comparison with literature cluster parameters} 
\begin{figure*}
    \centering
    \includegraphics[width=0.95\linewidth]{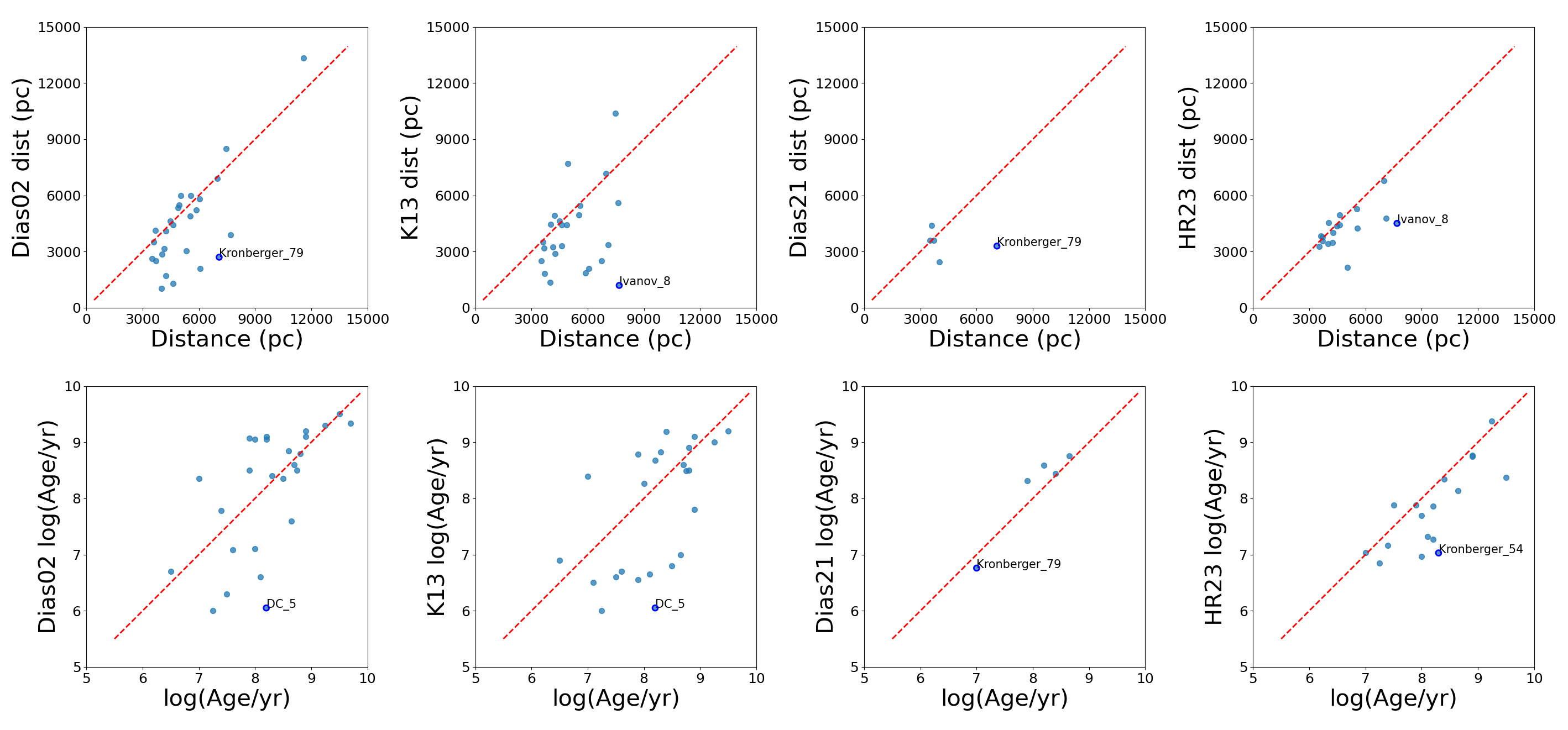}
    \caption{Comparison between the cluster parameters derived in this work (x-axes) and those reported in the literature catalogues (y-axes; Dias02, K13, Dias21, and HR23). 
    From top to bottom, the panels show comparisons of distance and age (log($\mathrm{Age/yr}$)). 
    Each point represents an individual cluster, with selected cluster names labelled for reference, and the red dashed line in each panel denotes the one-to-one relation.}
    \label{fig:xmatch_catlog}
\end{figure*}

Figure~\ref{fig:xmatch_catlog} presents a comparison between the cluster parameters derived in this work and those reported in several widely used OC catalogues: Dias02, K13, Dias21, and HR23. The top and bottom rows show comparisons of distance and $\log(\mathrm{Age/yr})$, respectively, with the red dashed line in each panel indicating the one-to-one relation. The distances adopted in this work are derived from the inverse of the weighted mean parallax of the member stars, where the zero-point correction described in Section~\ref{Gaia} has been applied and the weights are determined by the parallax uncertainties. The scatter seen in these comparisons likely reflects differences in membership selection, photometric data, and isochrone fitting strategies among the various catalogues.

Figure~\ref{fig:xmatch_catlog} also highlights the most discrepant outliers relative to the literature values. Regarding distances, Kronberger~79 and Ivanov~8 were previously classified as nearby or intermediate-distance clusters ($d < 3$~kpc), whereas our analysis places both clusters beyond $\sim$4~kpc. This discrepancy may reflect the small number of members identified in earlier studies ($N < 60$), which can lead to biased distance estimates due to insufficient sampling and stronger field-star contamination.

For clusters with $\log(\mathrm{Age/yr}) \gtrsim 8.30$, the ages derived here are in general agreement with literature values, with most systems lying close to the one-to-one relation. At younger ages ($\log(\mathrm{Age/yr}) < 8.30$), a larger scatter is seen among different catalogues, with notable examples including DC~5, Kronberger~79, and Kronberger~54. Several clusters previously classified as very young tend to have moderately older ages in this work, though the offsets vary between catalogues. This increased scatter likely reflects the intrinsic difficulty of age-dating young clusters, where the main-sequence turn-off is poorly defined and age constraints rely on pre-main-sequence features or sparse upper main-sequence stars. The large extinction values reported for some of these clusters ($A_{\rm v} > 3.8$~mag in HR23) may further contribute to the age uncertainties. Isochrone-based ages for young clusters should therefore be treated with caution and ideally complemented by additional constraints such as spectroscopic or kinematic data.

\begin{figure*}
    \centering
    \includegraphics[width=0.875\linewidth]{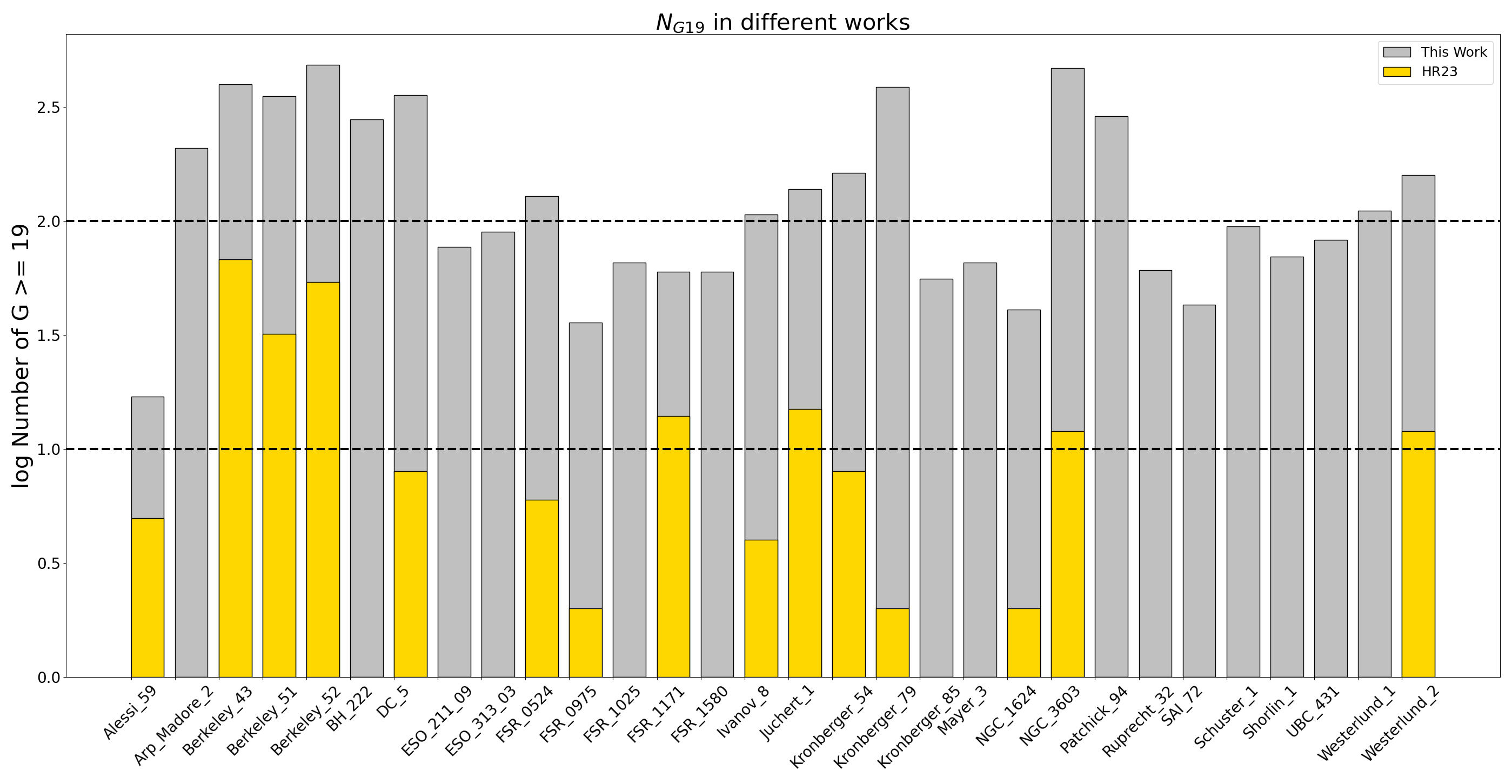}
    \caption{
    Comparison of faint-star members ($G \geq 19$~mag) between this work and HR23. 
    The grey bars represent the logarithmic number of faint-star members identified in this work, while the yellow bars correspond to those reported by HR23.
    }
    \label{fig:G19}
\end{figure*}

The number of faint members ($G \geq 19$~mag) reported in CG20 and Dias21 is generally small, with very few stars identified at the faint end. To further evaluate our method, we compare our faint-star counts with those from HR23 in Figure~\ref{fig:G19}. The grey bars show the logarithmic number of faint members recovered in this work, and the yellow bars the corresponding values from HR23. Our results systematically recover a substantially larger faint-star population, typically by about one order of magnitude.

Overall, the cluster parameters derived in this work are in good agreement with literature values, with most clusters lying close to the one-to-one relation. The remaining scatter can largely be attributed to differences in membership selection, photometric data, and fitting strategies among the various catalogues.

\subsection{Comparison with \texttt{pyUPMASK} and \texttt{Gaia\_oc\_amd}} 
To evaluate the quality of our membership determinations, we compare our results with those from two independent algorithms: \texttt{pyUPMASK} \citep{pyupmask} and \texttt{Gaia\_oc\_amd} \citep{VMG23}: 
\texttt{pyUPMASK} \footnote{\url{https://github.com/msolpera/pyUPMASK}} is a Python implementation of the UPMASK algorithm \citep{krone2014} that estimates membership probabilities through iterative clustering and spatial uniformity tests \citep[e.g.][]{He22a,Zhong22,Qin23}. We run it in five-dimensional astrometric space ($\mu_{\alpha}^{*}$, $\mu_{\delta}$, $\varpi$, $\alpha$, $\delta$) using the Mini-Batch K-Means clustering method; \texttt{Gaia\_oc\_amd} \footnote{\url{https://github.com/MGJvanGroeningen/Gaia\_oc\_amd}} \citep{VMG23} is a machine-learning pipeline developed for \textit{Gaia}~DR3. It uses a deep neural network trained on high-confidence members of known clusters to predict membership probabilities in an eight-dimensional astrometric and photometric parameter space.  Candidate members are selected by applying the same criteria as in this work: $r \leq r_{\rm clu}$ and $P \geq 0.5$.

For 28 of the 30 clusters in CG20 (excluding Arp~Madore~2 and FSR~1025), no prior estimates of $\log(\mathrm{Age/yr})$, $A_{\rm v}$, or distance are available, so we use the values from our isochrone fitting as initial conditions. Using \texttt{Gaia\_oc\_amd}, membership probabilities were successfully derived for 11 of the 30 clusters. The other 19 clusters were not processed further because their available member samples contained fewer than 15 stars, falling below the minimum sample size required for reliable model training. The results obtained with \texttt{pyUPMASK} for the 30 clusters are not fully satisfactory for some objects, as they show evidence of field-star contamination and poorly defined main sequences.

\begin{figure*}
    \centering
    \includegraphics[width=0.95\linewidth]{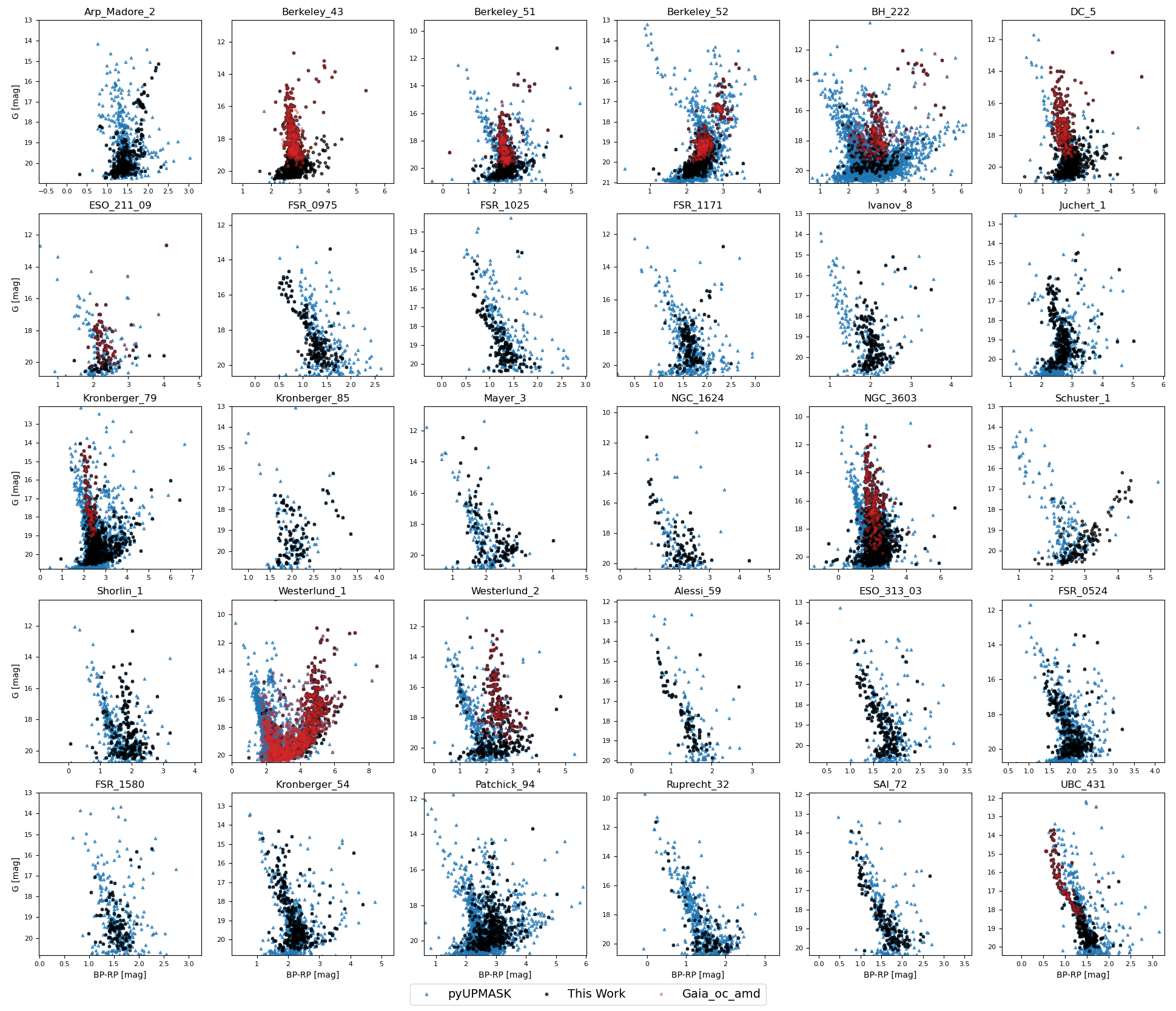}
    \caption{CMDs of the 30 clusters, showing a comparison between the member selections obtained with \texttt{pyUPMASK} \citep[][blue upward triangle]{pyupmask}, this work (black dots), and the \texttt{Gaia\_oc\_amd} \citep[][red pluses]{VMG23}. Each panel corresponds to one cluster and displays the distribution of stars in the $(G_{\rm BP}-G_{\rm RP})$ versus $G$ plane. The three methods generally recover a consistent main-sequence structure, while small differences in the faint or dispersed regions reveal the varying sensitivity of each method to low-probability or sparsely populated members.}
    \label{fig:othermethod}
\end{figure*}

Figure~\ref{fig:othermethod} shows the CMDs of all 30 known OCs from CG20 included in this work, comparing the member selections obtained with our method and \texttt{pyUPMASK}. The \texttt{Gaia\_oc\_amd} results are additionally shown for the 11 clusters for which this algorithm returned membership probabilities. These comparisons illustrate the differences in membership selection between our method and \texttt{pyUPMASK} for the full 30-cluster sample, while a three-way comparison is available for the 11 clusters with \texttt{Gaia\_oc\_amd} results. In several clusters, a larger number of faint, low-mass members is recovered, particularly in high-background regions. While the three methods agree on the overall main-sequence morphology, differences in the faint and dispersed regions reflect the varying sensitivity of each approach to low-probability members. Overall, our method provides a reasonable balance between completeness and reliability, which is particularly relevant for distant and low-contrast OCs.

\subsection{15 new OC candidates}
\begin{figure*}
    \centering
    \includegraphics[width=0.9\linewidth]{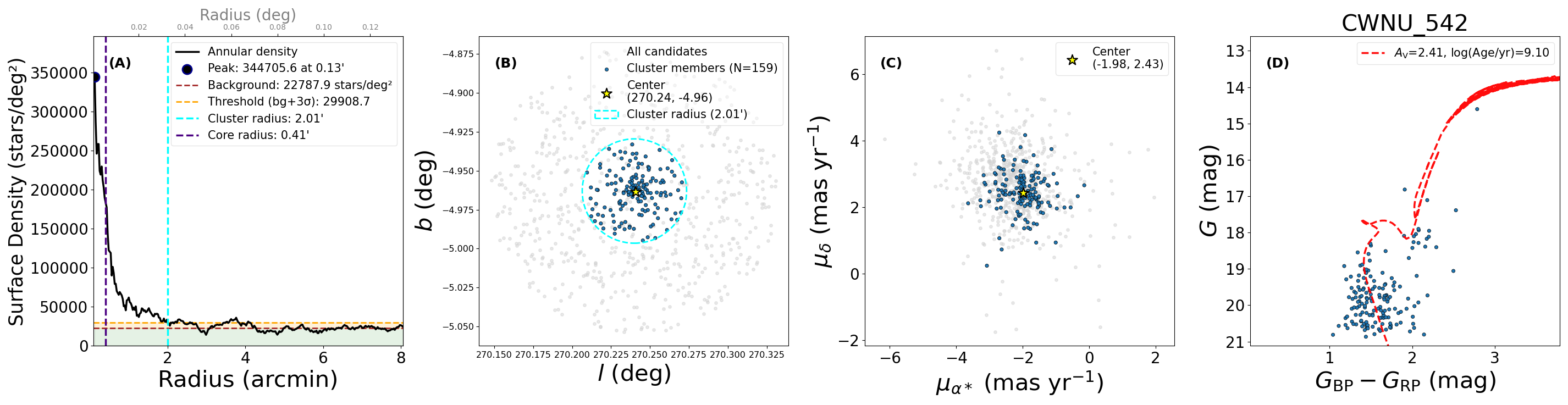}\\
    \includegraphics[width=0.9\linewidth]{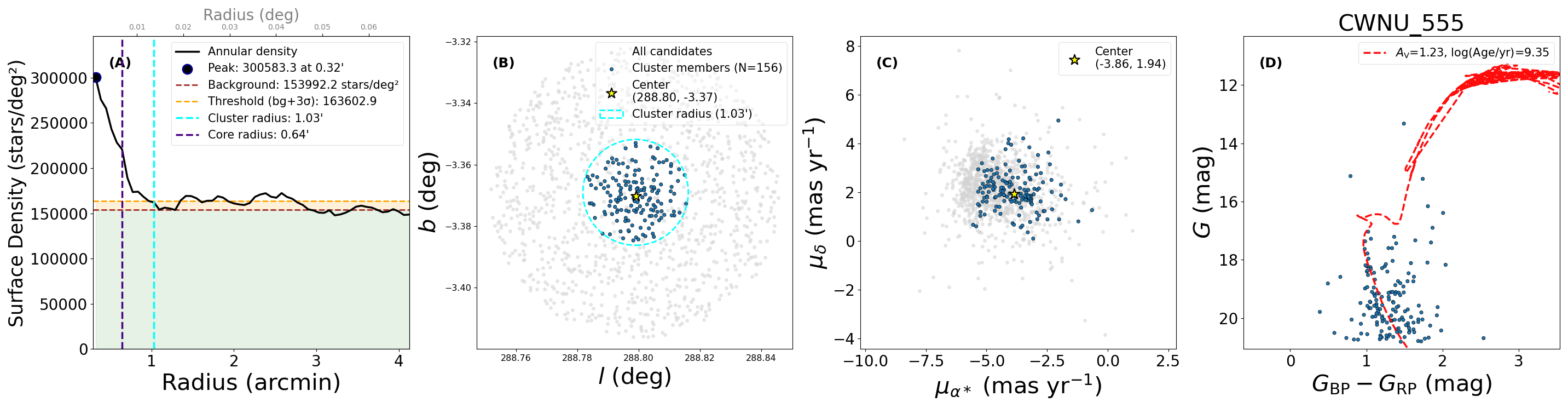}\\
    \caption{Same as Figure~\ref{fig:reidenti_oc}, but for new OC candidates CWNU\_542 (Type I) and CWNU\_555 (Type II). }
    \label{fig:cwnu542}
\end{figure*}

We revisited the research log of \citet{He23b} and performed membership determination and RDP analysis for a number of previously unconfirmed search regions. This yields 15 new cluster candidates, designated CWNU~542 through CWNU~556. Isochrone fitting to \textit{Gaia}~DR3 photometry, adopting the empirical metallicity distribution of the Galactic disc \citep{Hayden2015}, is performed to estimate fundamental parameters including age, extinction, and distance modulus. The resulting CMDs and RDPs are shown in Figure~\ref{fig:cwnu542}, and the derived parameters are listed in Table~\ref{tab:this_work}.

\subsection{Limitations of the membership identification} 
Despite the overall effectiveness of the adopted approach, several limitations should be noted. In the following sections, we discuss three major sources of limitation in detail: clusters with irregular morphology, field star contamination, and extreme extinction. Each of these factors can affect the accuracy of membership determinations and, consequently, the derived cluster parameters.

 \subsubsection{Clusters with irregular morphology} 
 The present work does not aim to characterise the detailed spatial morphology of individual clusters. Our analysis is designed to assess whether a statistically significant stellar overdensity is present, rather than to recover the intrinsic shape or internal substructure of each system. Clusters with irregular, elongated, or tidally distorted morphologies may therefore not be fully captured by the adopted RDP fitting, and their structural parameters should be interpreted with care.

\subsubsection{Field star contamination} \label{sec:field con}
The adopted method may be less effective for clusters with small physical sizes (e.g. $\lesssim 5$~pc) and fewer than $\sim$30 candidate members with $G<19$~mag. In such cases, the intrinsic cluster signal is weak, and the available astrometric and spatial information may be insufficient to separate the cluster reliably from fluctuations in the surrounding field population. The uncertainties in the RDP-based member selection are also sensitive to the level of interstellar extinction. In high-extinction, low-contrast environments, the observed stellar density is affected by spatially variable completeness, making it harder to define a clean cluster boundary. 

As a result, the estimated background density $\rho_{\rm bg}$ and its dispersion $\sigma_{\rm bg}$ depend on the adopted radial binning scheme in the concentric annuli. This bin dependence propagates into the threshold-based determination of the cluster boundary radius $r_{\rm clu}$, introducing an additional source of uncertainty in the spatial selection of candidate members. Such effects are expected to be more pronounced for distant clusters projected onto the Galactic plane, where extinction is strong and the cluster-to-field contrast is low.

\begin{figure*}
    \centering
    \includegraphics[width=0.9\linewidth]{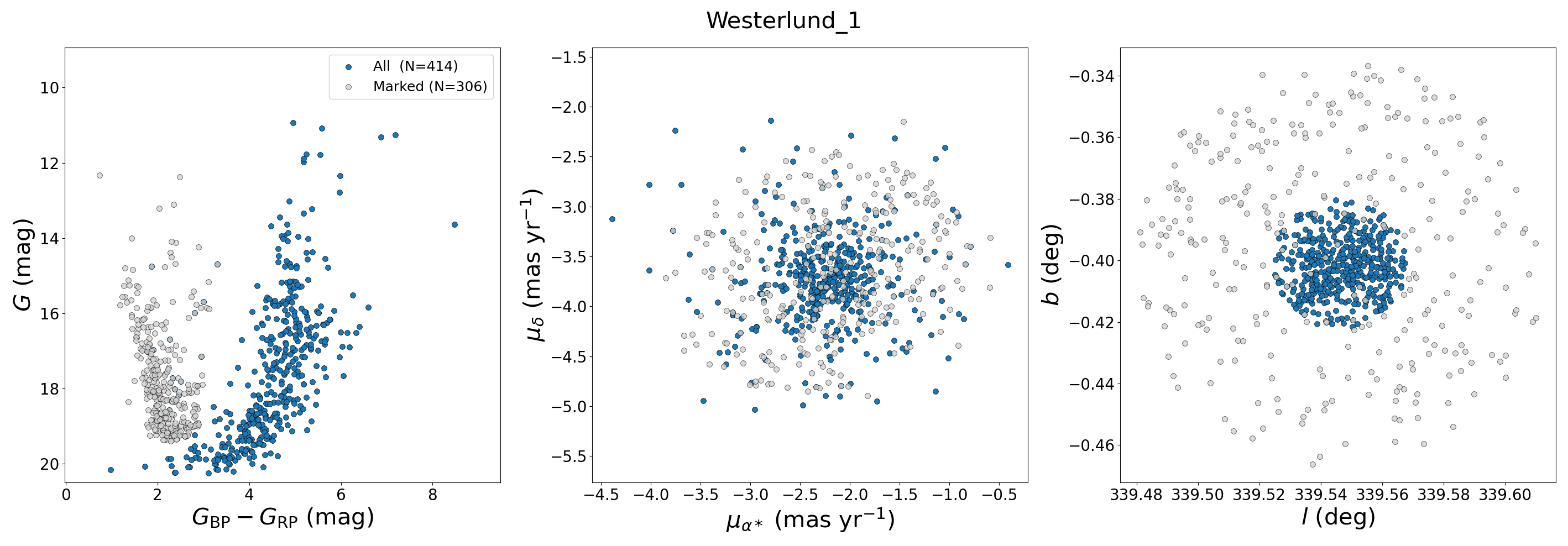} \\
    \includegraphics[width=0.9\linewidth]{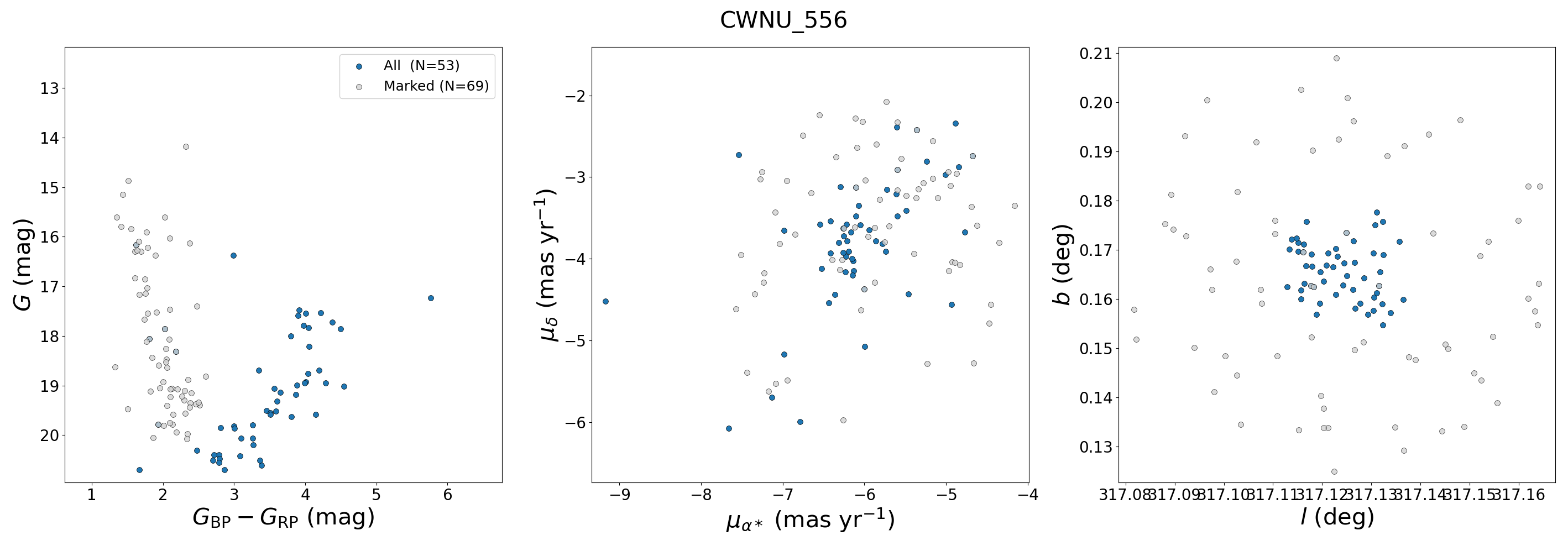} 
    
    \caption{Visual inspection of the candidate member distributions for Westerlund\_1 (top row) and CWNU\_556 (bottom row). For each cluster, the left panels show the CMD, the middle panels display the proper-motion diagram ($\mu_{\alpha*}$--$\mu_{\delta}$), and the right panels present the spatial distribution in Galactic coordinates ($l$--$b$). Blue points denote all stars selected as candidate members by our astrometric membership criteria. Grey points indicate sources that were manually marked as likely field contaminants within the same region. This comparison highlights the level of residual field-star contamination affecting the member selection, particularly for high extincted clusters located in the Galactic plane.
}
\label{fig:mark_region}
\end{figure*}

For two particular cases, Westerlund~1 and CWNU~556, we performed a visual inspection to remove obvious field contaminants based on photometric, spatial, and kinematic criteria. As shown in Figure~\ref{fig:mark_region}, the sources identified as likely field stars are distributed fairly uniformly across the field, with only a small fraction projected near the cluster core. In the proper-motion diagram, these contaminants are broadly distributed and do not form a compact kinematic overdensity. These examples illustrate that membership selection based solely on \textit{Gaia}~DR3 astrometry can suffer from residual field-star contamination, particularly for clusters with ambiguous morphological signatures, and that manual inspection is sometimes necessary in such cases.

\begin{figure*}
    \centering
    \includegraphics[width=0.9\linewidth]{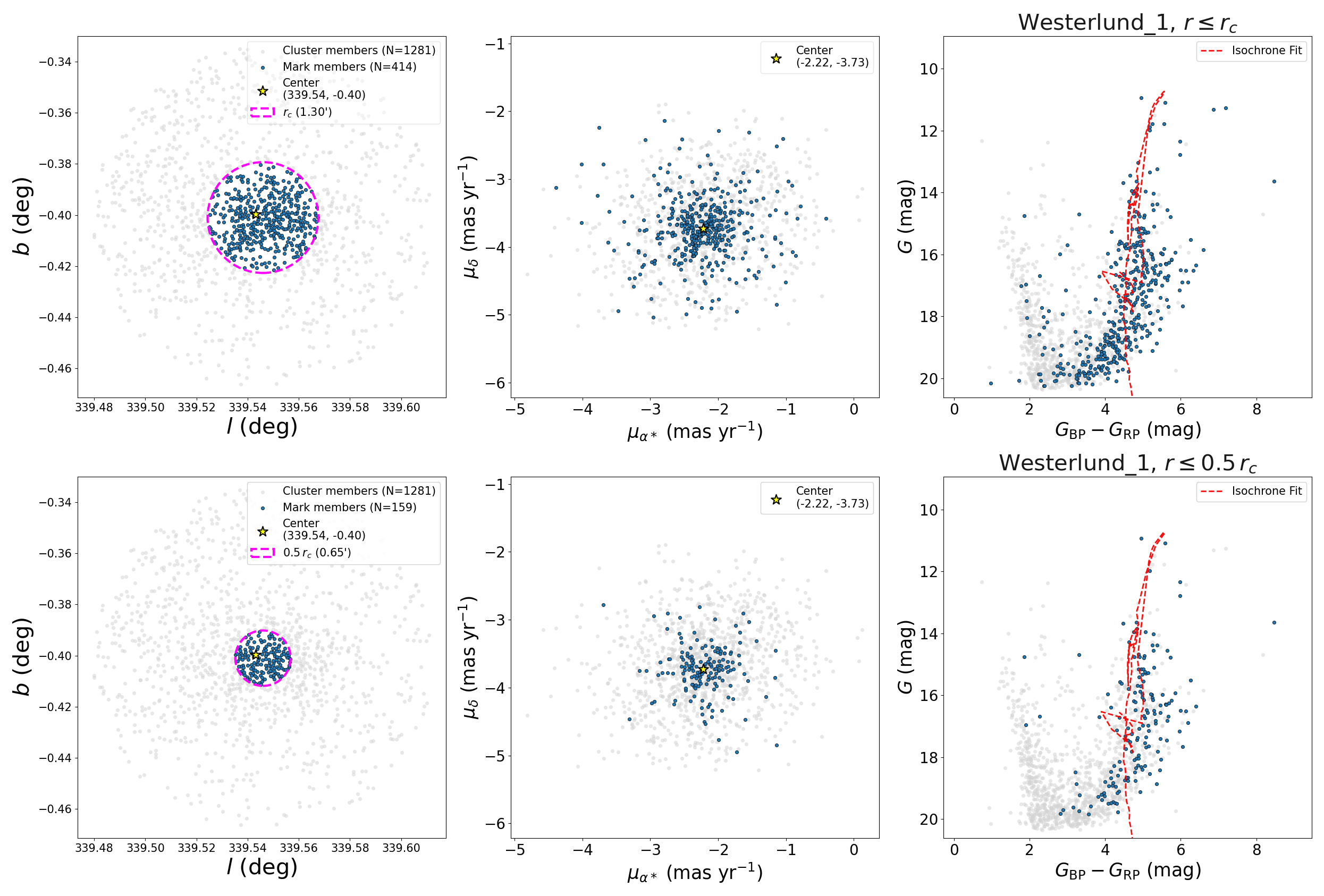} 
    \caption{Comparison of the spatial, kinematic, and photometric distributions of candidate members obtained under two radial selections. The left, middle, and right panels show the spatial distribution in Galactic coordinates ($l$--$b$), proper-motion distribution ($\mu_{\alpha*}$--$\mu_{\delta}$), and CMD respectively. The upper panels correspond to stars selected within the core radius ($r \leq r_{\rm c}$), while the lower panels show the results for a more restrictive selection of $r \leq 0.5 r_{\rm c}$. Blue points indicate stars satisfying the adopted membership criteria, and grey points denote the remaining sources in the same field.}

    \label{fig:diff_r_pick_w1}
\end{figure*}

For Westerlund~1 and CWNU~556, we therefore explored a reduced spatial selection to mitigate field contamination. Figure~\ref{fig:diff_r_pick_w1} shows the member distributions for Westerlund~1 under identical astrometric probability criteria but with different spatial cuts, $r \leq r_{\rm c}$ and $r \leq 0.5\,r_{\rm c}$. Selecting candidates within $r \leq r_{\rm c}$ already yields a noticeably cleaner main sequence and a more compact proper-motion distribution, consistent with a genuine stellar overdensity. The more restrictive cut of $r \leq 0.5\,r_{\rm c}$, however, leaves too few candidates for the cluster (e.g. Figure~\ref{fig:diff_r_pick_w1} bottom panel) to meaningfully assess its cluster nature. We therefore adopt $r \leq r_{\rm c}$ as the final spatial boundary for both clusters. These objects are marked with an asterisk in Table~\ref{tab:this_work}.

\subsubsection{Extreme extinction} \label{extre_a0}

Heavily extinguished clusters are predominantly located close to the Galactic plane and are typically young systems. In these cases, the CMD often lacks a well-defined main-sequence turn-off, which limits the accuracy of isochrone fitting. The derived ages and extinctions may therefore suffer from increased degeneracies, and metallicity-sensitive parameters inferred from photometric fitting alone are particularly poorly constrained. This limitation is intrinsic to photometric analyses of young, highly reddened clusters and cannot be fully overcome using \textit{Gaia}~DR3 data alone. Additional constraints from deeper photometric surveys or spectroscopic observations will be needed to improve the parameter determination for such systems. 

Eight clusters in our sample, namely Berkeley~43, BH~222, Patchick~94, Schuster~1, Westerlund~1, CWNU~545, CWNU~553, and CWNU~556, exhibit exceptionally high extinction ($A_{\rm v} \geq 6\,\mathrm{mag}$). These clusters are located at low Galactic latitudes and are affected by strong and spatially variable interstellar extinction, which substantially suppresses the detectability of their member stars in the \textit{Gaia} optical bands. Consequently, their sequences in the \textit{Gaia} CMDs are poorly defined, and the determination of fundamental parameters based solely on \textit{Gaia} photometry carries increased uncertainties.

\begin{figure*}
    \centering
    \includegraphics[width=0.9\linewidth]{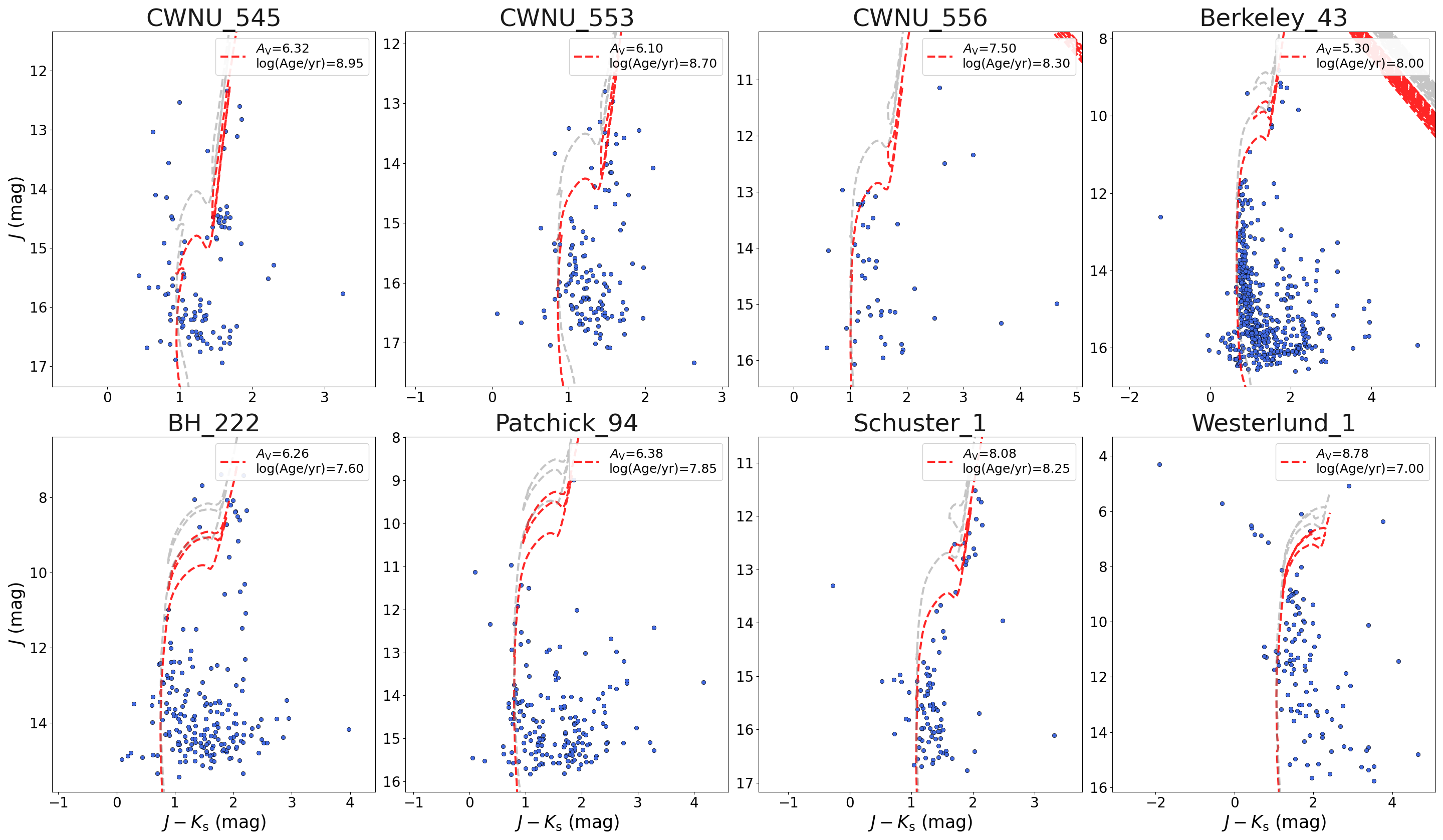}

    \caption{Near-infrared 2MASS CMDs of the eight high extinction clusters. Astrometrically selected candidate members are shown as blue points. The red and grey curves indicate the best-fitting theoretical isochrones for single-star and equal-mass binary populations, respectively. The use of 2MASS data significantly alleviates the effects of strong optical extinction and enables a clearer identification of the intrinsic cluster sequences.}

    \label{fig:iso_2mass}
\end{figure*}

To better constrain the physical properties of these highly reddened clusters, we complement the \textit{Gaia} astrometry with near-infrared photometry from the Two Micron All Sky Survey \citep[2MASS;][]{Skrutskie2006}, which provides uniform photometry in three bands ($J$ at 1.25\,$\mu$m, $H$ at 1.65\,$\mu$m, and $K_{\rm s}$ at 2.17\,$\mu$m). In the present analysis, only the $J$- and $K_{\rm s}$-band measurements are used to construct the $K_{\rm s}$ versus $(J-K_{\rm s})$ CMDs; the $H$-band photometry is not used. Compared with the \textit{Gaia} optical bands, the 2MASS $JHK_{\rm s}$ bands are significantly less affected by interstellar extinction and offer a clearer view of the intrinsic cluster sequences under heavy reddening conditions. We therefore construct near-infrared CMDs for these eight clusters using 2MASS photometry and perform isochrone fitting to investigate their stellar populations (Figure~\ref{fig:iso_2mass}). Candidate members selected on the basis of \textit{Gaia} astrometry are cross-matched with the 2MASS catalogue, yielding 1349 sources with valid near-infrared counterparts out of a total of 2666 candidate members across the eight clusters.

We further apply a polynomial extinction law to account for wavelength-dependent reddening. The adopted extinction coefficients for each photometric band are listed in Table~\ref{tab:extinction coefficient}. Isochrone fitting is then performed using 2MASS near-infrared photometry, yielding independent estimates of extinction and age for these clusters. This NIR-based analysis provides a consistency check on the \textit{Gaia}-derived results and helps to break degeneracies introduced by severe optical extinction. The resulting parameters are summarised in Table~\ref{tab:Gaia_2mass_age_a0}.

\begin{table*}[!htbp]
\centering
\caption{Comparison of cluster ages and extinctions derived from \textit{Gaia} and 2MASS CMD fitting.}
\label{tab:Gaia_2mass_age_a0}
\begin{threeparttable}
\renewcommand{\arraystretch}{1.4}
\setlength{\tabcolsep}{3.5pt}
\small
\begin{tabular}{lcccccccc}
\toprule
Cluster & $N_{\rm Gaia}$ & $N_{\rm 2MASS}$ & $\log(\mathrm{Age/yr})_{\rm Gaia}$ & $A_{\rm V,Gaia}$ & $\log(\mathrm{Age/yr})_{\rm 2MASS}$ & $A_{\rm V,2MASS}$ & $\log(\mathrm{Age/yr})$ & $A_{\rm V}$ \\
\midrule
CWNU\_545     & 158 & 104 & 9.15 & 6.48  & 8.95 & 6.32 & $9.05 \pm 0.10$ & $6.40 \pm 0.08$ \\
CWNU\_553     & 187 & 136 & 8.70 & 6.30  & 8.70 & 6.10 & $8.70 \pm 0.00$ & $6.20 \pm 0.10$ \\
CWNU\_556     & 53  & 46  & 8.00 & 10.00 & 8.30 & 7.50 & $8.15 \pm 0.15$ & $8.75 \pm 1.25$ \\
Berkeley\_43  & 721 & 482 & 7.90 & 6.30  & 8.00 & 5.30 & $7.95 \pm 0.05$ & $5.80 \pm 0.50$ \\
BH\_222       & 571 & 193 & 7.40 & 7.43  & 7.60 & 6.26 & $7.50 \pm 0.10$ & $6.85 \pm 0.59$ \\
Patchick\_94  & 434 & 177 & 7.90 & 6.50  & 7.85 & 6.38 & $7.88 \pm 0.03$ & $6.44 \pm 0.06$ \\
Schuster\_1   & 128 & 98  & 8.50 & 7.80  & 8.25 & 8.08 & $8.38 \pm 0.13$ & $7.94 \pm 0.14$ \\
Westerlund\_1 & 414 & 113 & 6.50 & 13.50 & 7.00 & 8.78 & $6.75 \pm 0.25$ & $11.14 \pm 2.36$ \\
\bottomrule
\end{tabular}
\begin{tablenotes}
\footnotesize
\item Notes. $N_{\rm Gaia}$ is the number of candidate members selected from the \textit{Gaia} astrometry, while $N_{\rm 2MASS}$ is the number of these stars with valid $J$- and $K_{\rm s}$-band photometry used in the near-infrared CMD fitting. The $H$-band photometry is not used. The final two columns list the adopted values of $\log(\mathrm{Age/yr})$ and $A_{\rm V}$, computed as the arithmetic means of the corresponding values derived from the \textit{Gaia} and 2MASS CMD fittings. The quoted uncertainties are taken as half of the absolute difference between the two estimates, that is, $\sigma=|X_{\rm Gaia}-X_{\rm 2MASS}|/2$.
\end{tablenotes}
\end{threeparttable}
\end{table*}

\section{Summary and conclusions}\label{sec:conclusions}

In this work, we present a probabilistic framework for the identification and first-order characterisation of distant OC candidates using \textit{Gaia}~DR3 astrometry. The method combines proper-motion and parallax-based membership probabilities with spatial density information and a refined RDP analysis, allowing us to evaluate cluster-like overdensities in distant and heavily contaminated Galactic fields. Rather than providing a definitive membership census, the approach offers a practical tool for assessing the presence of coherent stellar aggregates in low-contrast environments. 

We applied this framework to a sample of 30 distant OCs drawn from recent \textit{Gaia}-based catalogues, together with 15 new OC candidates identified here. Structural parameters were derived from the RDP analysis, and the systems were classified by their density contrast: objects with $\kappa > 3$ are labelled Type~I and the remainder Type~II. The derived ages, extinctions, and distances are broadly consistent with values reported in major compilations \citep{Dias02,Kharchenko13, Dias21, HR23}, with the largest discrepancies found for highly reddened clusters and for earlier catalogues based on limited member samples.

Several clusters in the sample exhibit exceptionally high extinction ($A_{\rm v} \gtrsim 6$\,mag), which significantly reduces the cluster-to-field contrast and increases the uncertainties in both the RDP-based boundary determination and the photometric parameter estimation. For such systems, the inferred cluster radius $r_{\rm clu}$ is sensitive to the adopted background definition and radial binning scheme. We therefore explored reduced spatial selections and, for Westerlund~1 and CWNU~556, adopted $r \leq r_{\rm c}$ to mitigate field contamination. For eight clusters with $A_{\rm v} \geq 6$\,mag (Berkeley~43, BH~222, Patchick~94, Schuster~1, Westerlund~1, CWNU~545, CWNU~553, and CWNU~556), we additionally performed isochrone fitting in the 2MASS $JHK_{\rm s}$ bands to obtain complementary age and extinction estimates. 

The framework presented here provides an efficient approach for recovering faint cluster signatures in \textit{Gaia} data, particularly in low-contrast and high extinction regions. Future improvements will require deeper near-infrared photometry and spectroscopic follow-up to establish more reliable memberships in such environments. The forthcoming \textit{Gaia}~DR4 release, with its improved astrometric precision and larger stellar sample, will further enhance the identification and characterisation of distant OCs, contributing to a more complete picture of the Galactic disc cluster population.

\section*{Data availability}
The cluster parameters listed in Table~\ref{tab:this_work} and the corresponding member catalogues are available in electronic form at the CDS via anonymous ftp to cdsarc.u-strasbg.fr (130.79.128.5) or via \url{http://cdsweb.u-strasbg.fr/cgi-bin/qcat?J/A+A/}.
The figures for all 45 OCs, corresponding to Fig.~\ref{fig:reidenti_oc}, are available at Zenodo: \url{https://doi.org/10.5281/zenodo.22240669}.

\begin{acknowledgements}
      This work was supported by National Natural Science Foundation of China through grants 12573023, the Fundamental Research Funds of China West Normal University (23kc011, 25KE031), and Project of the Natural Science Foundation of Sichuan Province (Grant No. 2026YFTX0019). This work has made use of data from the European Space Agency (ESA) mission \textit{Gaia} (\url{https://www.cosmos.esa.int/Gaia}), processed by the \textit{Gaia} Data Processing and Analysis Consortium (DPAC, \url{https://www.cosmos.esa.int/web/Gaia/dpac/consortium}). Funding for the DPAC has been provided by national institutions, in particular the institutions participating in the \textit{Gaia} Multilateral Agreement. 
\end{acknowledgements}

\bibliographystyle{aa}  
\bibliography{article}  

@ARTICLE{Bland-Hawthorn10,
       author = {{Bland-Hawthorn}, Joss and {Krumholz}, Mark R. and {Freeman}, Ken},
        title = "{The Long-term Evolution of the Galactic Disk Traced by Dissolving Star Clusters}",
      journal = {\apj},
         year = 2010,
        month = apr,
       volume = {713},
       number = {1},
        pages = {166-179},
          doi = {10.1088/0004-637X/713/1/166},
archivePrefix = {arXiv},
       eprint = {1002.4357},
 primaryClass = {astro-ph.GA},
       adsurl = {https://ui.adsabs.harvard.edu/abs/2010ApJ...713..166B}
}

@ARTICLE{Spina21,
       author = {{Spina}, L. and {Ting}, Y.-S. and {De Silva}, G.~M. and {Frankel}, N. and {Sharma}, S. and {Cantat-Gaudin}, T. and {Joyce}, M. and {Stello}, D. and {Karakas}, A.~I. and {Asplund}, M.~B. and {Nordlander}, T. and {Casagrande}, L. and {D'Orazi}, V. and {Casey}, A.~R. and {Cottrell}, P. and {Tepper-Garc{\'\i}a}, T. and {Baratella}, M. and {Kos}, J. and {{\v{C}}otar}, K. and {Bland-Hawthorn}, J. and {Buder}, S. and {Freeman}, K.~C. and {Hayden}, M.~R. and {Lewis}, G.~F. and {Lin}, J. and {Lind}, K. and {Martell}, S.~L. and {Schlesinger}, K.~J. and {Simpson}, J.~D. and {Zucker}, D.~B. and {Zwitter}, T.},
        title = "{The GALAH survey: tracing the Galactic disc with open clusters}",
      journal = {\mnras},
         year = 2021,
        month = may,
       volume = {503},
       number = {3},
        pages = {3279-3296},
          doi = {10.1093/mnras/stab471},
archivePrefix = {arXiv},
       eprint = {2011.02533},
 primaryClass = {astro-ph.GA},
       adsurl = {https://ui.adsabs.harvard.edu/abs/2021MNRAS.503.3279S}
}

@ARTICLE{Magrini17,
       author = {{Magrini}, L. and {Randich}, S. and {Kordopatis}, G. and {Prantzos}, N. and {Romano}, D. and {Chieffi}, A. and {Limongi}, M. and {Fran{\c{c}}ois}, P. and {Pancino}, E. and {Friel}, E. and {Bragaglia}, A. and {Tautvai{\v{s}}ien{\.{e}}}, G. and {Spina}, L. and {Overbeek}, J. and {Cantat-Gaudin}, T. and {Donati}, P. and {Vallenari}, A. and {Sordo}, R. and {Jim{\'e}nez-Esteban}, F.~M. and {Tang}, B. and {Drazdauskas}, A. and {Sousa}, S. and {Duffau}, S. and {Jofr{\'e}}, P. and {Gilmore}, G. and {Feltzing}, S. and {Alfaro}, E. and {Bensby}, T. and {Flaccomio}, E. and {Koposov}, S. and {Lanzafame}, A. and {Smiljanic}, R. and {Bayo}, A. and {Carraro}, G. and {Casey}, A.~R. and {Costado}, M.~T. and {Damiani}, F. and {Franciosini}, E. and {Hourihane}, A. and {Lardo}, C. and {Lewis}, J. and {Monaco}, L. and {Morbidelli}, L. and {Sacco}, G. and {Sbordone}, L. and {Worley}, C.~C. and {Zaggia}, S.},
        title = "{The Gaia-ESO Survey: radial distribution of abundances in the Galactic disc from open clusters and young-field stars}",
      journal = {\aap},
         year = 2017,
        month = jun,
       volume = {603},
          eid = {A2},
        pages = {A2},
          doi = {10.1051/0004-6361/201630294},
archivePrefix = {arXiv},
       eprint = {1703.00762},
 primaryClass = {astro-ph.GA},
       adsurl = {https://ui.adsabs.harvard.edu/abs/2017A&A...603A...2M}
}

@ARTICLE{Lada03,
       author = {{Lada}, Charles J. and {Lada}, Elizabeth A.},
        title = "{Embedded Clusters in Molecular Clouds}",
      journal = {\araa},
         year = 2003,
        month = jan,
       volume = {41},
        pages = {57-115},
          doi = {10.1146/annurev.astro.41.011802.094844},
archivePrefix = {arXiv},
       eprint = {astro-ph/0301540},
 primaryClass = {astro-ph},
       adsurl = {https://ui.adsabs.harvard.edu/abs/2003ARA&A..41...57L}
}

@ARTICLE{Friel1995,
        title = "{The Old Open Clusters Of The Milky Way}",
        author = {{Friel}, E.~D.},
      journal = {\araa},
         year = 1995,
        month = jan,
       volume = {33},
        pages = {381-414},
          doi = {10.1146/annurev.aa.33.090195.002121},
       adsurl = {https://ui.adsabs.harvard.edu/abs/1995ARA&A..33..381F}
}

@ARTICLE{JA1982,
       author = {{Janes}, K. and {Adler}, D.},
        title = "{Open clusters and galactic structure.}",
      journal = {\apjs},
         year = 1982,
        month = jul,
       volume = {49},
        pages = {425-446},
          doi = {10.1086/190805},
       adsurl = {https://ui.adsabs.harvard.edu/abs/1982ApJS...49..425J}
}

@ARTICLE{Soubiran2018,
       author = {{Soubiran}, C. and {Cantat-Gaudin}, T. and {Romero-G{\'o}mez}, M. and {Casamiquela}, L. and {Jordi}, C. and {Vallenari}, A. and {Antoja}, T. and {Balaguer-N{\'u}{\~n}ez}, L. and {Bossini}, D. and {Bragaglia}, A. and {Carrera}, R. and {Castro-Ginard}, A. and {Figueras}, F. and {Heiter}, U. and {Katz}, D. and {Krone-Martins}, A. and {Le Campion}, J.-F. and {Moitinho}, A. and {Sordo}, R.},
        title = "{Open cluster kinematics with Gaia DR2}",
      journal = {\aap},
         year = 2018,
        month = nov,
       volume = {619},
          eid = {A155},
        pages = {A155},
          doi = {10.1051/0004-6361/201834020},
archivePrefix = {arXiv},
       eprint = {1808.01613},
 primaryClass = {astro-ph.SR},
       adsurl = {https://ui.adsabs.harvard.edu/abs/2018A&A...619A.155S}
}

@ARTICLE{CG20,
       author = {{Cantat-Gaudin}, T. and {Anders}, F. and {Castro-Ginard}, A. and {Jordi}, C. and {Romero-G{\'o}mez}, M. and {Soubiran}, C. and {Casamiquela}, L. and {Tarricq}, Y. and {Moitinho}, A. and {Vallenari}, A. and {Bragaglia}, A. and {Krone-Martins}, A. and {Kounkel}, M.},
        title = "{Painting a portrait of the Galactic disc with its stellar clusters}",
      journal = {\aap},
         year = 2020,
        month = aug,
       volume = {640},
          eid = {A1},
        pages = {A1},
          doi = {10.1051/0004-6361/202038192},
archivePrefix = {arXiv},
       eprint = {2004.07274},
 primaryClass = {astro-ph.GA},
       adsurl = {https://ui.adsabs.harvard.edu/abs/2020A&A...640A...1C}
}

@ARTICLE{he23c,
       author = {{He}, Zhihong},
        title = "{Geometry and Kinematics of a Dancing Milky Way: Unveiling the Precession and Inclination Variation across the Galactic Plane via Open Clusters}",
      journal = {\apjl},
         year = 2023,
        month = sep,
       volume = {954},
       number = {1},
          eid = {L9},
        pages = {L9},
          doi = {10.3847/2041-8213/ace77d},
archivePrefix = {arXiv},
       eprint = {2306.17545},
 primaryClass = {astro-ph.GA},
       adsurl = {https://ui.adsabs.harvard.edu/abs/2023ApJ...954L...9H}
}

@ARTICLE{Brown21,
       author = {{Gaia Collaboration} and {Brown}, A.~G.~A. and {Vallenari}, A. and {Prusti}, T. and {de Bruijne}, J.~H.~J. and {Babusiaux}, C. and {Biermann}, M. and {Creevey}, O.~L. and {Evans}, D.~W. and {Eyer}, L. and {Hutton}, A. and {Jansen}, F. and {Jordi}, C. and {Klioner}, S.~A. and {Lammers}, U. and {Lindegren}, L. and {Luri}, X. and {Mignard}, F. and {Panem}, C. and {Pourbaix}, D. and {Randich}, S. and {Sartoretti}, P. and {Soubiran}, C. and {Walton}, N.~A. and {Arenou}, F. and {Bailer-Jones}, C.~A.~L. and {Bastian}, U. and {Cropper}, M. and {Drimmel}, R. and {Katz}, D. and {Lattanzi}, M.~G. and {van Leeuwen}, F. and {Bakker}, J. and {Cacciari}, C. and {Casta{\~n}eda}, J. and {De Angeli}, F. and {Ducourant}, C. and {Fabricius}, C. and {Fouesneau}, M. and {Fr{\'e}mat}, Y. and {Guerra}, R. and {Guerrier}, A. and {Guiraud}, J. and {Jean-Antoine Piccolo}, A. and {Masana}, E. and {Messineo}, R. and {Mowlavi}, N. and {Nicolas}, C. and {Nienartowicz}, K. and {Pailler}, F. and {Panuzzo}, P. and {Riclet}, F. and {Roux}, W. and {Seabroke}, G.~M. and {Sordo}, R. and {Tanga}, P. and {Th{\'e}venin}, F. and {Gracia-Abril}, G. and {Portell}, J. and {Teyssier}, D. and {Altmann}, M. and {Andrae}, R. and {Bellas-Velidis}, I. and {Benson}, K. and {Berthier}, J. and {Blomme}, R. and {Brugaletta}, E. and {Burgess}, P.~W. and {Busso}, G. and {Carry}, B. and {Cellino}, A. and {Cheek}, N. and {Clementini}, G. and {Damerdji}, Y. and {Davidson}, M. and {Delchambre}, L. and {Dell'Oro}, A. and {Fern{\'a}ndez-Hern{\'a}ndez}, J. and {Galluccio}, L. and {Garc{\'\i}a-Lario}, P. and {Garcia-Reinaldos}, M. and {Gonz{\'a}lez-N{\'u}{\~n}ez}, J. and {Gosset}, E. and {Haigron}, R. and {Halbwachs}, J.-L. and {Hambly}, N.~C. and {Harrison}, D.~L. and {Hatzidimitriou}, D. and {Heiter}, U. and {Hern{\'a}ndez}, J. and {Hestroffer}, D. and {Hodgkin}, S.~T. and {Holl}, B. and {Jan{\ss}en}, K. and {Jevardat de Fombelle}, G. and {Jordan}, S. and {Krone-Martins}, A. and {Lanzafame}, A.~C. and {L{\"o}ffler}, W. and {Lorca}, A. and {Manteiga}, M. and {Marchal}, O. and {Marrese}, P.~M. and {Moitinho}, A. and {Mora}, A. and {Muinonen}, K. and {Osborne}, P. and {Pancino}, E. and {Pauwels}, T. and {Petit}, J.-M. and {Recio-Blanco}, A. and {Richards}, P.~J. and {Riello}, M. and {Rimoldini}, L. and {Robin}, A.~C. and {Roegiers}, T. and {Rybizki}, J. and {Sarro}, L.~M. and {Siopis}, C. and {Smith}, M. and {Sozzetti}, A. and {Ulla}, A. and {Utrilla}, E. and {van Leeuwen}, M. and {van Reeven}, W. and {Abbas}, U. and {Abreu Aramburu}, A. and {Accart}, S. and {Aerts}, C. and {Aguado}, J.~J. and {Ajaj}, M. and {Altavilla}, G. and {{\'A}lvarez}, M.~A. and {{\'A}lvarez Cid-Fuentes}, J. and {Alves}, J. and {Anderson}, R.~I. and {Anglada Varela}, E. and {Antoja}, T. and {Audard}, M. and {Baines}, D. and {Baker}, S.~G. and {Balaguer-N{\'u}{\~n}ez}, L. and {Balbinot}, E. and {Balog}, Z. and {Barache}, C. and {Barbato}, D. and {Barros}, M. and {Barstow}, M.~A. and {Bartolom{\'e}}, S. and {Bassilana}, J.-L. and {Bauchet}, N. and {Baudesson-Stella}, A. and {Becciani}, U. and {Bellazzini}, M. and {Bernet}, M. and {Bertone}, S. and {Bianchi}, L. and {Blanco-Cuaresma}, S. and {Boch}, T. and {Bombrun}, A. and {Bossini}, D. and {Bouquillon}, S. and {Bragaglia}, A. and {Bramante}, L. and {Breedt}, E. and {Bressan}, A. and {Brouillet}, N. and {Bucciarelli}, B. and {Burlacu}, A. and {Busonero}, D. and {Butkevich}, A.~G. and {Buzzi}, R. and {Caffau}, E. and {Cancelliere}, R. and {C{\'a}novas}, H. and {Cantat-Gaudin}, T. and {Carballo}, R. and {Carlucci}, T. and {Carnerero}, M.~I. and {Carrasco}, J.~M. and {Casamiquela}, L. and {Castellani}, M. and {Castro-Ginard}, A. and {Castro Sampol}, P. and {Chaoul}, L. and {Charlot}, P. and {Chemin}, L. and {Chiavassa}, A. and {Cioni}, M.-R.~L. and {Comoretto}, G. and {Cooper}, W.~J. and {Cornez}, T. and {Cowell}, S. and {Crifo}, F. and {Crosta}, M. and {Crowley}, C. and {Dafonte}, C. and {Dapergolas}, A. and {David}, M. and {David}, P.},
        title = "{Gaia Early Data Release 3. Summary of the contents and survey properties}",
      journal = {\aap},
         year = 2021,
        month = may,
       volume = {649},
          eid = {A1},
        pages = {A1},
          doi = {10.1051/0004-6361/202039657},
archivePrefix = {arXiv},
       eprint = {2012.01533},
 primaryClass = {astro-ph.GA},
       adsurl = {https://ui.adsabs.harvard.edu/abs/2021A&A...649A...1G}
}

@ARTICLE{GAIADR2,
       author = {{Gaia Collaboration} and {Brown}, A.~G.~A. and {Vallenari}, A. and {Prusti}, T. and {de Bruijne}, J.~H.~J. and {Babusiaux}, C. and {Bailer-Jones}, C.~A.~L. and {Biermann}, M. and {Evans}, D.~W. and {Eyer}, L. and {Jansen}, F. and {Jordi}, C. and {Klioner}, S.~A. and {Lammers}, U. and {Lindegren}, L. and {Luri}, X. and {Mignard}, F. and {Panem}, C. and {Pourbaix}, D. and {Randich}, S. and {Sartoretti}, P. and {Siddiqui}, H.~I. and {Soubiran}, C. and {van Leeuwen}, F. and {Walton}, N.~A. and {Arenou}, F. and {Bastian}, U. and {Cropper}, M. and {Drimmel}, R. and {Katz}, D. and {Lattanzi}, M.~G. and {Bakker}, J. and {Cacciari}, C. and {Casta{\~n}eda}, J. and {Chaoul}, L. and {Cheek}, N. and {De Angeli}, F. and {Fabricius}, C. and {Guerra}, R. and {Holl}, B. and {Masana}, E. and {Messineo}, R. and {Mowlavi}, N. and {Nienartowicz}, K. and {Panuzzo}, P. and {Portell}, J. and {Riello}, M. and {Seabroke}, G.~M. and {Tanga}, P. and {Th{\'e}venin}, F. and {Gracia-Abril}, G. and {Comoretto}, G. and {Garcia-Reinaldos}, M. and {Teyssier}, D. and {Altmann}, M. and {Andrae}, R. and {Audard}, M. and {Bellas-Velidis}, I. and {Benson}, K. and {Berthier}, J. and {Blomme}, R. and {Burgess}, P. and {Busso}, G. and {Carry}, B. and {Cellino}, A. and {Clementini}, G. and {Clotet}, M. and {Creevey}, O. and {Davidson}, M. and {De Ridder}, J. and {Delchambre}, L. and {Dell'Oro}, A. and {Ducourant}, C. and {Fern{\'a}ndez-Hern{\'a}ndez}, J. and {Fouesneau}, M. and {Fr{\'e}mat}, Y. and {Galluccio}, L. and {Garc{\'\i}a-Torres}, M. and {Gonz{\'a}lez-N{\'u}{\~n}ez}, J. and {Gonz{\'a}lez-Vidal}, J.~J. and {Gosset}, E. and {Guy}, L.~P. and {Halbwachs}, J. -L. and {Hambly}, N.~C. and {Harrison}, D.~L. and {Hern{\'a}ndez}, J. and {Hestroffer}, D. and {Hodgkin}, S.~T. and {Hutton}, A. and {Jasniewicz}, G. and {Jean-Antoine-Piccolo}, A. and {Jordan}, S. and {Korn}, A.~J. and {Krone-Martins}, A. and {Lanzafame}, A.~C. and {Lebzelter}, T. and {L{\"o}ffler}, W. and {Manteiga}, M. and {Marrese}, P.~M. and {Mart{\'\i}n-Fleitas}, J.~M. and {Moitinho}, A. and {Mora}, A. and {Muinonen}, K. and {Osinde}, J. and {Pancino}, E. and {Pauwels}, T. and {Petit}, J. -M. and {Recio-Blanco}, A. and {Richards}, P.~J. and {Rimoldini}, L. and {Robin}, A.~C. and {Sarro}, L.~M. and {Siopis}, C. and {Smith}, M. and {Sozzetti}, A. and {S{\"u}veges}, M. and {Torra}, J. and {van Reeven}, W. and {Abbas}, U. and {Abreu Aramburu}, A. and {Accart}, S. and {Aerts}, C. and {Altavilla}, G. and {{\'A}lvarez}, M.~A. and {Alvarez}, R. and {Alves}, J. and {Anderson}, R.~I. and {Andrei}, A.~H. and {Anglada Varela}, E. and {Antiche}, E. and {Antoja}, T. and {Arcay}, B. and {Astraatmadja}, T.~L. and {Bach}, N. and {Baker}, S.~G. and {Balaguer-N{\'u}{\~n}ez}, L. and {Balm}, P. and {Barache}, C. and {Barata}, C. and {Barbato}, D. and {Barblan}, F. and {Barklem}, P.~S. and {Barrado}, D. and {Barros}, M. and {Barstow}, M.~A. and {Bartholom{\'e} Mu{\~n}oz}, S. and {Bassilana}, J. -L. and {Becciani}, U. and {Bellazzini}, M. and {Berihuete}, A. and {Bertone}, S. and {Bianchi}, L. and {Bienaym{\'e}}, O. and {Blanco-Cuaresma}, S. and {Boch}, T. and {Boeche}, C. and {Bombrun}, A. and {Borrachero}, R. and {Bossini}, D. and {Bouquillon}, S. and {Bourda}, G. and {Bragaglia}, A. and {Bramante}, L. and {Breddels}, M.~A. and {Bressan}, A. and {Brouillet}, N. and {Br{\"u}semeister}, T. and {Brugaletta}, E. and {Bucciarelli}, B. and {Burlacu}, A. and {Busonero}, D. and {Butkevich}, A.~G. and {Buzzi}, R. and {Caffau}, E. and {Cancelliere}, R. and {Cannizzaro}, G. and {Cantat-Gaudin}, T. and {Carballo}, R. and {Carlucci}, T. and {Carrasco}, J.~M. and {Casamiquela}, L. and {Castellani}, M. and {Castro-Ginard}, A. and {Charlot}, P. and {Chemin}, L. and {Chiavassa}, A. and {Cocozza}, G. and {Costigan}, G. and {Cowell}, S. and {Crifo}, F. and {Crosta}, M. and {Crowley}, C. and {Cuypers}, J. and {Dafonte}, C. and {Damerdji}, Y. and {Dapergolas}, A. and {David}, P. and {David}, M. and {de Laverny}, P. and {De Luise}, F. and {De March}, R. and {de Martino}, D. and {de Souza}, R. and {de Torres}, A. and {Debosscher}, J. and {del Pozo}, E. and {Delbo}, M. and {Delgado}, A. and {Delgado}, H.~E. and {Di Matteo}, P. and {Diakite}, S. and {Diener}, C. and {Distefano}, E. and {Dolding}, C. and {Drazinos}, P. and {Dur{\'a}n}, J. and {Edvardsson}, B. and {Enke}, H. and {Eriksson}, K. and {Esquej}, P. and {Eynard Bontemps}, G. and {Fabre}, C. and {Fabrizio}, M. and {Faigler}, S. and {Falc{\~a}o}, A.~J. and {Farr{\`a}s Casas}, M. and {Federici}, L. and {Fedorets}, G. and {Fernique}, P. and {Figueras}, F. and {Filippi}, F. and {Findeisen}, K. and {Fonti}, A. and {Fraile}, E. and {Fraser}, M. and {Fr{\'e}zouls}, B. and {Gai}, M. and {Galleti}, S. and {Garabato}, D. and {Garc{\'\i}a-Sedano}, F. and {Garofalo}, A. and {Garralda}, N. and {Gavel}, A. and {Gavras}, P. and {Gerssen}, J. and {Geyer}, R. and {Giacobbe}, P. and {Gilmore}, G. and {Girona}, S. and {Giuffrida}, G. and {Glass}, F. and {Gomes}, M. and {Granvik}, M. and {Gueguen}, A. and {Guerrier}, A. and {Guiraud}, J. and {Guti{\'e}rrez-S{\'a}nchez}, R. and {Haigron}, R. and {Hatzidimitriou}, D. and {Hauser}, M. and {Haywood}, M. and {Heiter}, U. and {Helmi}, A. and {Heu}, J. and {Hilger}, T. and {Hobbs}, D. and {Hofmann}, W. and {Holland}, G. and {Huckle}, H.~E. and {Hypki}, A. and {Icardi}, V. and {Jan{\ss}en}, K. and {Jevardat de Fombelle}, G. and {Jonker}, P.~G. and {Juh{\'a}sz}, {\'A}. L. and {Julbe}, F. and {Karampelas}, A. and {Kewley}, A. and {Klar}, J. and {Kochoska}, A. and {Kohley}, R. and {Kolenberg}, K. and {Kontizas}, M. and {Kontizas}, E. and {Koposov}, S.~E. and {Kordopatis}, G. and {Kostrzewa-Rutkowska}, Z. and {Koubsky}, P. and {Lambert}, S. and {Lanza}, A.~F. and {Lasne}, Y. and {Lavigne}, J. -B. and {Le Fustec}, Y. and {Le Poncin-Lafitte}, C. and {Lebreton}, Y. and {Leccia}, S. and {Leclerc}, N. and {Lecoeur-Taibi}, I. and {Lenhardt}, H. and {Leroux}, F. and {Liao}, S. and {Licata}, E. and {Lindstr{\o}m}, H.~E.~P. and {Lister}, T.~A. and {Livanou}, E. and {Lobel}, A. and {L{\'o}pez}, M. and {Managau}, S. and {Mann}, R.~G. and {Mantelet}, G. and {Marchal}, O. and {Marchant}, J.~M. and {Marconi}, M. and {Marinoni}, S. and {Marschalk{\'o}}, G. and {Marshall}, D.~J. and {Martino}, M. and {Marton}, G. and {Mary}, N. and {Massari}, D. and {Matijevi{\v{c}}}, G. and {Mazeh}, T. and {McMillan}, P.~J. and {Messina}, S. and {Michalik}, D. and {Millar}, N.~R. and {Molina}, D. and {Molinaro}, R. and {Moln{\'a}r}, L. and {Montegriffo}, P. and {Mor}, R. and {Morbidelli}, R. and {Morel}, T. and {Morris}, D. and {Mulone}, A.~F. and {Muraveva}, T. and {Musella}, I. and {Nelemans}, G. and {Nicastro}, L. and {Noval}, L. and {O'Mullane}, W. and {Ord{\'e}novic}, C. and {Ord{\'o}{\~n}ez-Blanco}, D. and {Osborne}, P. and {Pagani}, C. and {Pagano}, I. and {Pailler}, F. and {Palacin}, H. and {Palaversa}, L. and {Panahi}, A. and {Pawlak}, M. and {Piersimoni}, A.~M. and {Pineau}, F. -X. and {Plachy}, E. and {Plum}, G. and {Poggio}, E. and {Poujoulet}, E. and {Pr{\v{s}}a}, A. and {Pulone}, L. and {Racero}, E. and {Ragaini}, S. and {Rambaux}, N. and {Ramos-Lerate}, M. and {Regibo}, S. and {Reyl{\'e}}, C. and {Riclet}, F. and {Ripepi}, V. and {Riva}, A. and {Rivard}, A. and {Rixon}, G. and {Roegiers}, T. and {Roelens}, M. and {Romero-G{\'o}mez}, M. and {Rowell}, N. and {Royer}, F. and {Ruiz-Dern}, L. and {Sadowski}, G. and {Sagrist{\`a} Sell{\'e}s}, T. and {Sahlmann}, J. and {Salgado}, J. and {Salguero}, E. and {Sanna}, N. and {Santana-Ros}, T. and {Sarasso}, M. and {Savietto}, H. and {Schultheis}, M. and {Sciacca}, E. and {Segol}, M. and {Segovia}, J.~C. and {S{\'e}gransan}, D. and {Shih}, I. -C. and {Siltala}, L. and {Silva}, A.~F. and {Smart}, R.~L. and {Smith}, K.~W. and {Solano}, E. and {Solitro}, F. and {Sordo}, R. and {Soria Nieto}, S. and {Souchay}, J. and {Spagna}, A. and {Spoto}, F. and {Stampa}, U. and {Steele}, I.~A. and {Steidelm{\"u}ller}, H. and {Stephenson}, C.~A. and {Stoev}, H. and {Suess}, F.~F. and {Surdej}, J. and {Szabados}, L. and {Szegedi-Elek}, E. and {Tapiador}, D. and {Taris}, F. and {Tauran}, G. and {Taylor}, M.~B. and {Teixeira}, R. and {Terrett}, D. and {Teyssandier}, P. and {Thuillot}, W. and {Titarenko}, A. and {Torra Clotet}, F. and {Turon}, C. and {Ulla}, A. and {Utrilla}, E. and {Uzzi}, S. and {Vaillant}, M. and {Valentini}, G. and {Valette}, V. and {van Elteren}, A. and {Van Hemelryck}, E. and {van Leeuwen}, M. and {Vaschetto}, M. and {Vecchiato}, A. and {Veljanoski}, J. and {Viala}, Y. and {Vicente}, D. and {Vogt}, S. and {von Essen}, C. and {Voss}, H. and {Votruba}, V. and {Voutsinas}, S. and {Walmsley}, G. and {Weiler}, M. and {Wertz}, O. and {Wevers}, T. and {Wyrzykowski}, {\L}. and {Yoldas}, A. and {{\v{Z}}erjal}, M. and {Ziaeepour}, H. and {Zorec}, J. and {Zschocke}, S. and {Zucker}, S. and {Zurbach}, C. and {Zwitter}, T.},
        title = "{Gaia Data Release 2. Summary of the contents and survey properties}",
      journal = {\aap},
         year = 2018,
        month = aug,
       volume = {616},
          eid = {A1},
        pages = {A1},
          doi = {10.1051/0004-6361/201833051},
archivePrefix = {arXiv},
       eprint = {1804.09365},
 primaryClass = {astro-ph.GA},
       adsurl = {https://ui.adsabs.harvard.edu/abs/2018A&A...616A...1G}
}

@ARTICLE{GAIADR3,
       author = {{Gaia Collaboration} and {Vallenari}, A. and {Brown}, A.~G.~A. and {Prusti}, T. and {de Bruijne}, J.~H.~J. and {Arenou}, F. and {Babusiaux}, C. and {Biermann}, M. and {Creevey}, O.~L. and {Ducourant}, C. and {Evans}, D.~W. and {Eyer}, L. and {Guerra}, R. and {Hutton}, A. and {Jordi}, C. and {Klioner}, S.~A. and {Lammers}, U.~L. and {Lindegren}, L. and {Luri}, X. and {Mignard}, F. and {Panem}, C. and {Pourbaix}, D. and {Randich}, S. and {Sartoretti}, P. and {Soubiran}, C. and {Tanga}, P. and {Walton}, N.~A. and {Bailer-Jones}, C.~A.~L. and {Bastian}, U. and {Drimmel}, R. and {Jansen}, F. and {Katz}, D. and {Lattanzi}, M.~G. and {van Leeuwen}, F. and {Bakker}, J. and {Cacciari}, C. and {Casta{\~n}eda}, J. and {De Angeli}, F. and {Fabricius}, C. and {Fouesneau}, M. and {Fr{\'e}mat}, Y. and {Galluccio}, L. and {Guerrier}, A. and {Heiter}, U. and {Masana}, E. and {Messineo}, R. and {Mowlavi}, N. and {Nicolas}, C. and {Nienartowicz}, K. and {Pailler}, F. and {Panuzzo}, P. and {Riclet}, F. and {Roux}, W. and {Seabroke}, G.~M. and {Sordo}, R. and {Th{\'e}venin}, F. and {Gracia-Abril}, G. and {Portell}, J. and {Teyssier}, D. and {Altmann}, M. and {Andrae}, R. and {Audard}, M. and {Bellas-Velidis}, I. and {Benson}, K. and {Berthier}, J. and {Blomme}, R. and {Burgess}, P.~W. and {Busonero}, D. and {Busso}, G. and {C{\'a}novas}, H. and {Carry}, B. and {Cellino}, A. and {Cheek}, N. and {Clementini}, G. and {Damerdji}, Y. and {Davidson}, M. and {de Teodoro}, P. and {Nu{\~n}ez Campos}, M. and {Delchambre}, L. and {Dell'Oro}, A. and {Esquej}, P. and {Fern{\'a}ndez-Hern{\'a}ndez}, J. and {Fraile}, E. and {Garabato}, D. and {Garc{\'\i}a-Lario}, P. and {Gosset}, E. and {Haigron}, R. and {Halbwachs}, J. -L. and {Hambly}, N.~C. and {Harrison}, D.~L. and {Hern{\'a}ndez}, J. and {Hestroffer}, D. and {Hodgkin}, S.~T. and {Holl}, B. and {Jan{\ss}en}, K. and {Jevardat de Fombelle}, G. and {Jordan}, S. and {Krone-Martins}, A. and {Lanzafame}, A.~C. and {L{\"o}ffler}, W. and {Marchal}, O. and {Marrese}, P.~M. and {Moitinho}, A. and {Muinonen}, K. and {Osborne}, P. and {Pancino}, E. and {Pauwels}, T. and {Recio-Blanco}, A. and {Reyl{\'e}}, C. and {Riello}, M. and {Rimoldini}, L. and {Roegiers}, T. and {Rybizki}, J. and {Sarro}, L.~M. and {Siopis}, C. and {Smith}, M. and {Sozzetti}, A. and {Utrilla}, E. and {van Leeuwen}, M. and {Abbas}, U. and {{\'A}brah{\'a}m}, P. and {Abreu Aramburu}, A. and {Aerts}, C. and {Aguado}, J.~J. and {Ajaj}, M. and {Aldea-Montero}, F. and {Altavilla}, G. and {{\'A}lvarez}, M.~A. and {Alves}, J. and {Anders}, F. and {Anderson}, R.~I. and {Anglada Varela}, E. and {Antoja}, T. and {Baines}, D. and {Baker}, S.~G. and {Balaguer-N{\'u}{\~n}ez}, L. and {Balbinot}, E. and {Balog}, Z. and {Barache}, C. and {Barbato}, D. and {Barros}, M. and {Barstow}, M.~A. and {Bartolom{\'e}}, S. and {Bassilana}, J. -L. and {Bauchet}, N. and {Becciani}, U. and {Bellazzini}, M. and {Berihuete}, A. and {Bernet}, M. and {Bertone}, S. and {Bianchi}, L. and {Binnenfeld}, A. and {Blanco-Cuaresma}, S. and {Blazere}, A. and {Boch}, T. and {Bombrun}, A. and {Bossini}, D. and {Bouquillon}, S. and {Bragaglia}, A. and {Bramante}, L. and {Breedt}, E. and {Bressan}, A. and {Brouillet}, N. and {Brugaletta}, E. and {Bucciarelli}, B. and {Burlacu}, A. and {Butkevich}, A.~G. and {Buzzi}, R. and {Caffau}, E. and {Cancelliere}, R. and {Cantat-Gaudin}, T. and {Carballo}, R. and {Carlucci}, T. and {Carnerero}, M.~I. and {Carrasco}, J.~M. and {Casamiquela}, L. and {Castellani}, M. and {Castro-Ginard}, A. and {Chaoul}, L. and {Charlot}, P. and {Chemin}, L. and {Chiaramida}, V. and {Chiavassa}, A. and {Chornay}, N. and {Comoretto}, G. and {Contursi}, G. and {Cooper}, W.~J. and {Cornez}, T. and {Cowell}, S. and {Crifo}, F. and {Cropper}, M. and {Crosta}, M. and {Crowley}, C. and {Dafonte}, C. and {Dapergolas}, A. and {David}, M. and {David}, P. and {de Laverny}, P. and {De Luise}, F. and {De March}, R. and {De Ridder}, J. and {de Souza}, R. and {de Torres}, A. and {del Peloso}, E.~F. and {del Pozo}, E. and {Delbo}, M. and {Delgado}, A. and {Delisle}, J. -B. and {Demouchy}, C. and {Dharmawardena}, T.~E. and {Di Matteo}, P. and {Diakite}, S. and {Diener}, C. and {Distefano}, E. and {Dolding}, C. and {Edvardsson}, B. and {Enke}, H. and {Fabre}, C. and {Fabrizio}, M. and {Faigler}, S. and {Fedorets}, G. and {Fernique}, P. and {Fienga}, A. and {Figueras}, F. and {Fournier}, Y. and {Fouron}, C. and {Fragkoudi}, F. and {Gai}, M. and {Garcia-Gutierrez}, A. and {Garcia-Reinaldos}, M. and {Garc{\'\i}a-Torres}, M. and {Garofalo}, A. and {Gavel}, A. and {Gavras}, P. and {Gerlach}, E. and {Geyer}, R. and {Giacobbe}, P. and {Gilmore}, G. and {Girona}, S. and {Giuffrida}, G. and {Gomel}, R. and {Gomez}, A. and {Gonz{\'a}lez-N{\'u}{\~n}ez}, J. and {Gonz{\'a}lez-Santamar{\'\i}a}, I. and {Gonz{\'a}lez-Vidal}, J.~J. and {Granvik}, M. and {Guillout}, P. and {Guiraud}, J. and {Guti{\'e}rrez-S{\'a}nchez}, R. and {Guy}, L.~P. and {Hatzidimitriou}, D. and {Hauser}, M. and {Haywood}, M. and {Helmer}, A. and {Helmi}, A. and {Sarmiento}, M.~H. and {Hidalgo}, S.~L. and {Hilger}, T. and {H{\l}adczuk}, N. and {Hobbs}, D. and {Holland}, G. and {Huckle}, H.~E. and {Jardine}, K. and {Jasniewicz}, G. and {Jean-Antoine Piccolo}, A. and {Jim{\'e}nez-Arranz}, {\'O}. and {Jorissen}, A. and {Juaristi Campillo}, J. and {Julbe}, F. and {Karbevska}, L. and {Kervella}, P. and {Khanna}, S. and {Kontizas}, M. and {Kordopatis}, G. and {Korn}, A.~J. and {K{\'o}sp{\'a}l}, {\'A}. and {Kostrzewa-Rutkowska}, Z. and {Kruszy{\'n}ska}, K. and {Kun}, M. and {Laizeau}, P. and {Lambert}, S. and {Lanza}, A.~F. and {Lasne}, Y. and {Le Campion}, J. -F. and {Lebreton}, Y. and {Lebzelter}, T. and {Leccia}, S. and {Leclerc}, N. and {Lecoeur-Taibi}, I. and {Liao}, S. and {Licata}, E.~L. and {Lindstr{\o}m}, H.~E.~P. and {Lister}, T.~A. and {Livanou}, E. and {Lobel}, A. and {Lorca}, A. and {Loup}, C. and {Madrero Pardo}, P. and {Magdaleno Romeo}, A. and {Managau}, S. and {Mann}, R.~G. and {Manteiga}, M. and {Marchant}, J.~M. and {Marconi}, M. and {Marcos}, J. and {Marcos Santos}, M.~M.~S. and {Mar{\'\i}n Pina}, D. and {Marinoni}, S. and {Marocco}, F. and {Marshall}, D.~J. and {Martin Polo}, L. and {Mart{\'\i}n-Fleitas}, J.~M. and {Marton}, G. and {Mary}, N. and {Masip}, A. and {Massari}, D. and {Mastrobuono-Battisti}, A. and {Mazeh}, T. and {McMillan}, P.~J. and {Messina}, S. and {Michalik}, D. and {Millar}, N.~R. and {Mints}, A. and {Molina}, D. and {Molinaro}, R. and {Moln{\'a}r}, L. and {Monari}, G. and {Mongui{\'o}}, M. and {Montegriffo}, P. and {Montero}, A. and {Mor}, R. and {Mora}, A. and {Morbidelli}, R. and {Morel}, T. and {Morris}, D. and {Muraveva}, T. and {Murphy}, C.~P. and {Musella}, I. and {Nagy}, Z. and {Noval}, L. and {Oca{\~n}a}, F. and {Ogden}, A. and {Ordenovic}, C. and {Osinde}, J.~O. and {Pagani}, C. and {Pagano}, I. and {Palaversa}, L. and {Palicio}, P.~A. and {Pallas-Quintela}, L. and {Panahi}, A. and {Payne-Wardenaar}, S. and {Pe{\~n}alosa Esteller}, X. and {Penttil{\"a}}, A. and {Pichon}, B. and {Piersimoni}, A.~M. and {Pineau}, F. -X. and {Plachy}, E. and {Plum}, G. and {Poggio}, E. and {Pr{\v{s}}a}, A. and {Pulone}, L. and {Racero}, E. and {Ragaini}, S. and {Rainer}, M. and {Raiteri}, C.~M. and {Rambaux}, N. and {Ramos}, P. and {Ramos-Lerate}, M. and {Re Fiorentin}, P. and {Regibo}, S. and {Richards}, P.~J. and {Rios Diaz}, C. and {Ripepi}, V. and {Riva}, A. and {Rix}, H. -W. and {Rixon}, G. and {Robichon}, N. and {Robin}, A.~C. and {Robin}, C. and {Roelens}, M. and {Rogues}, H.~R.~O. and {Rohrbasser}, L. and {Romero-G{\'o}mez}, M. and {Rowell}, N. and {Royer}, F. and {Ruz Mieres}, D. and {Rybicki}, K.~A. and {Sadowski}, G. and {S{\'a}ez N{\'u}{\~n}ez}, A. and {Sagrist{\`a} Sell{\'e}s}, A. and {Sahlmann}, J. and {Salguero}, E. and {Samaras}, N. and {Sanchez Gimenez}, V. and {Sanna}, N. and {Santove{\~n}a}, R. and {Sarasso}, M. and {Schultheis}, M. and {Sciacca}, E. and {Segol}, M. and {Segovia}, J.~C. and {S{\'e}gransan}, D. and {Semeux}, D. and {Shahaf}, S. and {Siddiqui}, H.~I. and {Siebert}, A. and {Siltala}, L. and {Silvelo}, A. and {Slezak}, E. and {Slezak}, I. and {Smart}, R.~L. and {Snaith}, O.~N. and {Solano}, E. and {Solitro}, F. and {Souami}, D. and {Souchay}, J. and {Spagna}, A. and {Spina}, L. and {Spoto}, F. and {Steele}, I.~A. and {Steidelm{\"u}ller}, H. and {Stephenson}, C.~A. and {S{\"u}veges}, M. and {Surdej}, J. and {Szabados}, L. and {Szegedi-Elek}, E. and {Taris}, F. and {Taylor}, M.~B. and {Teixeira}, R. and {Tolomei}, L. and {Tonello}, N. and {Torra}, F. and {Torra}, J. and {Torralba Elipe}, G. and {Trabucchi}, M. and {Tsounis}, A.~T. and {Turon}, C. and {Ulla}, A. and {Unger}, N. and {Vaillant}, M.~V. and {van Dillen}, E. and {van Reeven}, W. and {Vanel}, O. and {Vecchiato}, A. and {Viala}, Y. and {Vicente}, D. and {Voutsinas}, S. and {Weiler}, M. and {Wevers}, T. and {Wyrzykowski}, {\L}. and {Yoldas}, A. and {Yvard}, P. and {Zhao}, H. and {Zorec}, J. and {Zucker}, S. and {Zwitter}, T.},
        title = "{Gaia Data Release 3. Summary of the content and survey properties}",
      journal = {\aap},
         year = 2023,
        month = jun,
       volume = {674},
          eid = {A1},
        pages = {A1},
          doi = {10.1051/0004-6361/202243940},
archivePrefix = {arXiv},
       eprint = {2208.00211},
 primaryClass = {astro-ph.GA},
       adsurl = {https://ui.adsabs.harvard.edu/abs/2023A&A...674A...1G}
}

@ARTICLE{Bossini2019,
       author = {{Bossini}, D. and {Vallenari}, A. and {Bragaglia}, A. and {Cantat-Gaudin}, T. and {Sordo}, R. and {Balaguer-N{\'u}{\~n}ez}, L. and {Jordi}, C. and {Moitinho}, A. and {Soubiran}, C. and {Casamiquela}, L. and {Carrera}, R. and {Heiter}, U.},
        title = "{Age determination for 269 Gaia DR2 open clusters}",
      journal = {\aap},
         year = 2019,
        month = mar,
       volume = {623},
          eid = {A108},
        pages = {A108},
          doi = {10.1051/0004-6361/201834693},
archivePrefix = {arXiv},
       eprint = {1901.04733},
 primaryClass = {astro-ph.SR},
       adsurl = {https://ui.adsabs.harvard.edu/abs/2019A&A...623A.108B}
}

@ARTICLE{Dias25,
       author = {{Dias}, W.~S. and {Monteiro}, H.},
        title = "{Discovery of 28 Open Clusters with Gaia DR3}",
      journal = {\rmxaa},
         year = 2025,
        month = apr,
       volume = {61},
        pages = {3-13},
          doi = {10.22201/ia.01851101p.2025.61.01.01},
       adsurl = {https://ui.adsabs.harvard.edu/abs/2025RMxAA..61a...3D}
}

@ARTICLE{Dias21,
       author = {{Dias}, W.~S. and {Monteiro}, H. and {Moitinho}, A. and {L{\'e}pine}, J.~R.~D. and {Carraro}, G. and {Paunzen}, E. and {Alessi}, B. and {Villela}, L.},
        title = "{Updated parameters of 1743 open clusters based on Gaia DR2}",
      journal = {\mnras},
         year = 2021,
        month = jun,
       volume = {504},
       number = {1},
        pages = {356-371},
          doi = {10.1093/mnras/stab770},
archivePrefix = {arXiv},
       eprint = {2103.12829},
 primaryClass = {astro-ph.SR},
       adsurl = {https://ui.adsabs.harvard.edu/abs/2021MNRAS.504..356D}
}

@ARTICLE{Cavallo24,
       author = {{Cavallo}, Lorenzo and {Spina}, Lorenzo and {Carraro}, Giovanni and {Magrini}, Laura and {Poggio}, Eloisa and {Cantat-Gaudin}, Tristan and {Pasquato}, Mario and {Lucatello}, Sara and {Ortolani}, Sergio and {Schiappacasse-Ulloa}, Jose},
        title = "{Parameter Estimation for Open Clusters using an Artificial Neural Network with a QuadTree-based Feature Extractor}",
      journal = {\aj},
         year = 2024,
        month = jan,
       volume = {167},
       number = {1},
          eid = {12},
        pages = {12},
          doi = {10.3847/1538-3881/ad07e5},
archivePrefix = {arXiv},
       eprint = {2311.03009},
 primaryClass = {astro-ph.GA},
       adsurl = {https://ui.adsabs.harvard.edu/abs/2024AJ....167...12C}
}

@ARTICLE{he21a,
       author = {{He}, Zhi-Hong and {Xu}, Ye and {Hao}, Chao-Jie and {Wu}, Zhen-Yu and {Li}, Jing-Jing},
        title = "{A catalogue of 74 new open clusters found in Gaia Data-Release 2}",
      journal = {\raa},
         year = 2021,
        month = may,
       volume = {21},
       number = {4},
          eid = {093},
        pages = {093},
          doi = {10.1088/1674-4527/21/4/93},
archivePrefix = {arXiv},
       eprint = {2010.14870},
 primaryClass = {astro-ph.GA},
       adsurl = {https://ui.adsabs.harvard.edu/abs/2021RAA....21...93H}
}

@ARTICLE{he22b,
       author = {{He}, Zhihong and {Wang}, Kun and {Luo}, Yangping and {Li}, Jing and {Liu}, Xiaochen and {Jiang}, Qingquan},
        title = "{A Blind All-sky Search for Star Clusters in Gaia EDR3: 886 Clusters within 1.2 kpc of the Sun}",
      journal = {\apjs},
         year = 2022,
        month = sep,
       volume = {262},
       number = {1},
          eid = {7},
        pages = {7},
          doi = {10.3847/1538-4365/ac7c17},
archivePrefix = {arXiv},
       eprint = {2206.12170},
 primaryClass = {astro-ph.GA},
       adsurl = {https://ui.adsabs.harvard.edu/abs/2022ApJS..262....7H}
}

@ARTICLE{he22a,
       author = {{He}, Zhihong and {Li}, Chunyan and {Zhong}, Jing and {Liu}, Guimei and {Bai}, Leya and {Qin}, Songmei and {Jiang}, Yueyue and {Zhang}, Xi and {Chen}, Li},
        title = "{New Open-cluster Candidates Found in the Galactic Disk Using Gaia DR2/EDR3 Data}",
      journal = {\apjs},
         year = 2022,
        month = may,
       volume = {260},
       number = {1},
          eid = {8},
        pages = {8},
          doi = {10.3847/1538-4365/ac5cbb},
archivePrefix = {arXiv},
       eprint = {2203.05177},
 primaryClass = {astro-ph.GA},
       adsurl = {https://ui.adsabs.harvard.edu/abs/2022ApJS..260....8H}
}

@ARTICLE{He23b,
       author = {{He}, Zhihong and {Luo}, Yangping and {Wang}, Kun and {Ren}, Anbing and {Peng}, Liming and {Cui}, Qian and {Liu}, Xiaochen and {Jiang}, Qingquan},
        title = "{Survey for Distant Stellar Aggregates in the Galactic Disk: Detecting 2000 Star Clusters and Candidates, along with the Dwarf Galaxy IC 10}",
      journal = {\apjs},
         year = 2023,
        month = aug,
       volume = {267},
       number = {2},
          eid = {34},
        pages = {34},
          doi = {10.3847/1538-4365/acd6fa},
archivePrefix = {arXiv},
       eprint = {2305.10269},
 primaryClass = {astro-ph.GA},
       adsurl = {https://ui.adsabs.harvard.edu/abs/2023ApJS..267...34H}
}

@ARTICLE{Dias02,
       author = {{Dias}, W.~S. and {Alessi}, B.~S. and {Moitinho}, A. and {L{\'e}pine}, J.~R.~D.},
        title = "{New catalogue of optically visible open clusters and candidates}",
      journal = {\aap},
         year = 2002,
        month = jul,
       volume = {389},
        pages = {871-873},
          doi = {10.1051/0004-6361:20020668},
archivePrefix = {arXiv},
       eprint = {astro-ph/0203351},
 primaryClass = {astro-ph},
       adsurl = {https://ui.adsabs.harvard.edu/abs/2002A&A...389..871D}
}

@ARTICLE{Kharchenko13,
       author = {{Kharchenko}, N.~V. and {Piskunov}, A.~E. and {Schilbach}, E. and
         {R{\"o}ser}, S. and {Scholz}, R. -D.},
        title = "{Global survey of star clusters in the Milky Way. II. The catalogue of basic parameters}",
      journal = {\aap},
         year = 2013,
        month = oct,
       volume = {558},
          eid = {A53},
        pages = {A53},
          doi = {10.1051/0004-6361/201322302},
archivePrefix = {arXiv},
       eprint = {1308.5822},
 primaryClass = {astro-ph.GA},
       adsurl = {https://ui.adsabs.harvard.edu/abs/2013A&A...558A..53K}
}

@ARTICLE{CG18,
       author = {{Cantat-Gaudin}, T. and {Jordi}, C. and {Vallenari}, A. and
         {Bragaglia}, A. and {Balaguer-N{\'u}{\~n}ez}, L. and {Soubiran}, C. and
         {Bossini}, D. and {Moitinho}, A. and {Castro-Ginard}, A. and
         {Krone-Martins}, A. and {Casamiquela}, L. and {Sordo}, R. and
         {Carrera}, R.},
        title = "{A Gaia DR2 view of the open cluster population in the Milky Way}",
      journal = {\aap},
         year = 2018,
        month = oct,
       volume = {618},
          eid = {A93},
        pages = {A93},
          doi = {10.1051/0004-6361/201833476},
archivePrefix = {arXiv},
       eprint = {1805.08726},
 primaryClass = {astro-ph.GA},
       adsurl = {https://ui.adsabs.harvard.edu/abs/2018A&A...618A..93C}
}

@ARTICLE{Sim19,
       author = {{Sim}, Gyuheon and {Lee}, Sang Hyun and {Ann}, Hong Bae and
         {Kim}, Seunghyeon},
        title = "{207 New Open Star Clusters within 1 kpc from Gaia Data Release 2}",
      journal = {\jkas},
         year = 2019,
        month = oct,
       volume = {52},
        pages = {145-158},
          doi = {10.5303/JKAS.2019.52.5.145},
archivePrefix = {arXiv},
       eprint = {1907.06872},
 primaryClass = {astro-ph.SR},
       adsurl = {https://ui.adsabs.harvard.edu/abs/2019JKAS...52..145S}
}

@ARTICLE{LP19,
       author = {{Liu}, Lei and {Pang}, Xiaoying},
        title = "{A Catalog of Newly Identified Star Clusters in Gaia DR2}",
      journal = {\apjs},
         year = 2019,
        month = dec,
       volume = {245},
       number = {2},
          eid = {32},
        pages = {32},
          doi = {10.3847/1538-4365/ab530a},
archivePrefix = {arXiv},
       eprint = {1910.12600},
 primaryClass = {astro-ph.GA},
       adsurl = {https://ui.adsabs.harvard.edu/abs/2019ApJS..245...32L}
}

@ARTICLE{Castro22,
       author = {{Castro-Ginard}, A. and {Jordi}, C. and {Luri}, X. and {Cantat-Gaudin}, T. and {Carrasco}, J.~M. and {Casamiquela}, L. and {Anders}, F. and {Balaguer-N{\'u}{\~n}ez}, L. and {Badia}, R.~M.},
        title = "{Hunting for open clusters in Gaia EDR3: 628 new open clusters found with OCfinder}",
      journal = {\aap},
         year = 2022,
        month = may,
       volume = {661},
          eid = {A118},
        pages = {A118},
          doi = {10.1051/0004-6361/202142568},
archivePrefix = {arXiv},
       eprint = {2111.01819},
 primaryClass = {astro-ph.GA},
       adsurl = {https://ui.adsabs.harvard.edu/abs/2022A&A...661A.118C}
}

@ARTICLE{Castro20,
       author = {{Castro-Ginard}, A. and {Jordi}, C. and {Luri}, X. and
         {{\'A}lvarez Cid-Fuentes}, J. and {Casamiquela}, L. and {Anders}, F. and
         {Cantat-Gaudin}, T. and {Mongui{\'o}}, M. and
         {Balaguer-N{\'u}{\~n}ez}, L. and {Sol{\`a}}, S. and {Badia}, R.~M.},
        title = "{Hunting for open clusters in Gaia DR2: 582 new open clusters in the Galactic disc}",
      journal = {\aap},
         year = 2020,
        month = mar,
       volume = {635},
          eid = {A45},
        pages = {A45},
          doi = {10.1051/0004-6361/201937386},
archivePrefix = {arXiv},
       eprint = {2001.07122},
 primaryClass = {astro-ph.GA},
       adsurl = {https://ui.adsabs.harvard.edu/abs/2020A&A...635A..45C}
}

@ARTICLE{Ferreira21,
       author = {{Ferreira}, F.~A. and {Corradi}, W.~J.~B. and {Maia}, F.~F.~S. and {Angelo}, M.~S. and {Santos}, J.~F.~C., Jr.},
        title = "{New star clusters discovered towards the Galactic bulge direction using Gaia DR2}",
      journal = {\mnras},
         year = 2021,
        month = mar,
       volume = {502},
       number = {1},
        pages = {L90-L94},
          doi = {10.1093/mnrasl/slab011},
archivePrefix = {arXiv},
       eprint = {2101.10982},
 primaryClass = {astro-ph.GA},
       adsurl = {https://ui.adsabs.harvard.edu/abs/2021MNRAS.502L..90F}
}

@ARTICLE{Ferreira20,
       author = {{Ferreira}, F.~A. and {Corradi}, W.~J.~B. and {Maia}, F.~F.~S. and
         {Angelo}, M.~S. and {Santos}, J.~F.~C., Jr.},
        title = "{Discovery and astrophysical properties of Galactic open clusters in dense stellar fields using Gaia DR2}",
      journal = {\mnras},
         year = 2020,
        month = jun,
       volume = {496},
       number = {2},
        pages = {2021-2038},
          doi = {10.1093/mnras/staa1684},
archivePrefix = {arXiv},
       eprint = {2006.05611},
 primaryClass = {astro-ph.SR},
       adsurl = {https://ui.adsabs.harvard.edu/abs/2020MNRAS.496.2021F}
}

@ARTICLE{HR23,
       author = {{Hunt}, Emily L. and {Reffert}, Sabine},
        title = "{Improving the open cluster census. II. An all-sky cluster catalogue with Gaia DR3}",
      journal = {\aap},
         year = 2023,
        month = may,
       volume = {673},
          eid = {A114},
        pages = {A114},
          doi = {10.1051/0004-6361/202346285},
archivePrefix = {arXiv},
       eprint = {2303.13424},
 primaryClass = {astro-ph.GA},
       adsurl = {https://ui.adsabs.harvard.edu/abs/2023A&A...673A.114H}
}

@ARTICLE{Carraro07,
       author = {{Carraro}, G. and {Geisler}, D. and {Villanova}, S. and {Frinchaboy}, P.~M. and {Majewski}, S.~R.},
        title = "{Old open clusters in the outer Galactic disk}",
      journal = {\aap},
         year = 2007,
        month = dec,
       volume = {476},
       number = {1},
        pages = {217-227},
          doi = {10.1051/0004-6361:20078113},
archivePrefix = {arXiv},
       eprint = {0709.2126},
 primaryClass = {astro-ph},
       adsurl = {https://ui.adsabs.harvard.edu/abs/2007A&A...476..217C}
}

@ARTICLE{Castro21,
       author = {{Castro-Ginard}, A. and {McMillan}, P.~J. and {Luri}, X. and {Jordi}, C. and {Romero-G{\'o}mez}, M. and {Cantat-Gaudin}, T. and {Casamiquela}, L. and {Tarricq}, Y. and {Soubiran}, C. and {Anders}, F.},
        title = "{Milky Way spiral arms from open clusters in Gaia EDR3}",
      journal = {\aap},
         year = 2021,
        month = aug,
       volume = {652},
          eid = {A162},
        pages = {A162},
          doi = {10.1051/0004-6361/202039751},
archivePrefix = {arXiv},
       eprint = {2105.04590},
 primaryClass = {astro-ph.GA},
       adsurl = {https://ui.adsabs.harvard.edu/abs/2021A&A...652A.162C}
}

@ARTICLE{Perren22,
       author = {{Perren}, G.~I. and {Pera}, M.~S. and {Navone}, H.~D. and {V{\'a}zquez}, R.~A.},
        title = "{An analysis of the most distant cataloged open clusters. Re-assessing fundamental parameters with Gaia EDR3 and ASteCA}",
      journal = {\aap},
         year = 2022,
        month = jul,
       volume = {663},
          eid = {A131},
        pages = {A131},
          doi = {10.1051/0004-6361/202243288},
archivePrefix = {arXiv},
       eprint = {2204.02153},
 primaryClass = {astro-ph.GA},
       adsurl = {https://ui.adsabs.harvard.edu/abs/2022A&A...663A.131P}
}

@ARTICLE{Lindegren2021a,
       author = {{Lindegren}, L. and {Bastian}, U. and {Biermann}, M. and {Bombrun}, A. and {de Torres}, A. and {Gerlach}, E. and {Geyer}, R. and {Hern{\'a}ndez}, J. and {Hilger}, T. and {Hobbs}, D. and {Klioner}, S.~A. and {Lammers}, U. and {McMillan}, P.~J. and {Ramos-Lerate}, M. and {Steidelm{\"u}ller}, H. and {Stephenson}, C.~A. and {van Leeuwen}, F.},
        title = "{Gaia Early Data Release 3. Parallax bias versus magnitude, colour, and position}",
      journal = {\aap},
         year = 2021,
        month = may,
       volume = {649},
          eid = {A4},
        pages = {A4},
          doi = {10.1051/0004-6361/202039653},
archivePrefix = {arXiv},
       eprint = {2012.01742},
 primaryClass = {astro-ph.IM},
       adsurl = {https://ui.adsabs.harvard.edu/abs/2021A&A...649A...4L}
}

@ARTICLE{Lindegren2021b,
       author = {{Lindegren}, L. and {Klioner}, S.~A. and {Hern{\'a}ndez}, J. and {Bombrun}, A. and {Ramos-Lerate}, M. and {Steidelm{\"u}ller}, H. and {Bastian}, U. and {Biermann}, M. and {de Torres}, A. and {Gerlach}, E. and {Geyer}, R. and {Hilger}, T. and {Hobbs}, D. and {Lammers}, U. and {McMillan}, P.~J. and {Stephenson}, C.~A. and {Casta{\~n}eda}, J. and {Davidson}, M. and {Fabricius}, C. and {Gracia-Abril}, G. and {Portell}, J. and {Rowell}, N. and {Teyssier}, D. and {Torra}, F. and {Bartolom{\'e}}, S. and {Clotet}, M. and {Garralda}, N. and {Gonz{\'a}lez-Vidal}, J.~J. and {Torra}, J. and {Abbas}, U. and {Altmann}, M. and {Anglada Varela}, E. and {Balaguer-N{\'u}{\~n}ez}, L. and {Balog}, Z. and {Barache}, C. and {Becciani}, U. and {Bernet}, M. and {Bertone}, S. and {Bianchi}, L. and {Bouquillon}, S. and {Brown}, A.~G.~A. and {Bucciarelli}, B. and {Busonero}, D. and {Butkevich}, A.~G. and {Buzzi}, R. and {Cancelliere}, R. and {Carlucci}, T. and {Charlot}, P. and {Cioni}, M.-R.~L. and {Crosta}, M. and {Crowley}, C. and {del Peloso}, E.~F. and {del Pozo}, E. and {Drimmel}, R. and {Esquej}, P. and {Fienga}, A. and {Fraile}, E. and {Gai}, M. and {Garcia-Reinaldos}, M. and {Guerra}, R. and {Hambly}, N.~C. and {Hauser}, M. and {Jan{\ss}en}, K. and {Jordan}, S. and {Kostrzewa-Rutkowska}, Z. and {Lattanzi}, M.~G. and {Liao}, S. and {Licata}, E. and {Lister}, T.~A. and {L{\"o}ffler}, W. and {Marchant}, J.~M. and {Masip}, A. and {Mignard}, F. and {Mints}, A. and {Molina}, D. and {Mora}, A. and {Morbidelli}, R. and {Murphy}, C.~P. and {Pagani}, C. and {Panuzzo}, P. and {Pe{\~n}alosa Esteller}, X. and {Poggio}, E. and {Re Fiorentin}, P. and {Riva}, A. and {Sagrist{\`a} Sell{\'e}s}, A. and {Sanchez Gimenez}, V. and {Sarasso}, M. and {Sciacca}, E. and {Siddiqui}, H.~I. and {Smart}, R.~L. and {Souami}, D. and {Spagna}, A. and {Steele}, I.~A. and {Taris}, F. and {Utrilla}, E. and {van Reeven}, W. and {Vecchiato}, A.},
        title = "{Gaia Early Data Release 3. The astrometric solution}",
      journal = {\aap},
         year = 2021,
        month = may,
       volume = {649},
          eid = {A2},
        pages = {A2},
          doi = {10.1051/0004-6361/202039709},
archivePrefix = {arXiv},
       eprint = {2012.03380},
 primaryClass = {astro-ph.IM},
       adsurl = {https://ui.adsabs.harvard.edu/abs/2021A&A...649A...2L}
}

@ARTICLE{Vasilevskis1958,
       author = {{Vasilevskis}, S. and {Klemola}, A. and {Preston}, G.},
        title = "{Relative proper motions of stars in the region of the open cluster NGC 6633.}",
      journal = {\aj},
         year = 1958,
        month = jan,
       volume = {63},
        pages = {387-395},
          doi = {10.1086/107787},
       adsurl = {https://ui.adsabs.harvard.edu/abs/1958AJ.....63..387V}
}

@ARTICLE{Sander1971,
       author = {{Sanders}, W.~L.},
        title = "{An improved method for computing membership probabilities in open clusters.}",
      journal = {\aap},
         year = 1971,
        month = sep,
       volume = {14},
        pages = {226-232},
       adsurl = {https://ui.adsabs.harvard.edu/abs/1971A&A....14..226S}
}

@ARTICLE{Zhao1990,
       author = {{Zhao}, J.~L. and {He}, Y.~P.},
        title = "{An improved method for membership determination of stellar clusters with proper motions with different accuracies.}",
      journal = {\aap},
         year = 1990,
        month = oct,
       volume = {237},
        pages = {54},
       adsurl = {https://ui.adsabs.harvard.edu/abs/1990A&A...237...54Z}
}

@ARTICLE{cabrea_alfaro1990,
       author = {{Cabrera-Cano}, J. and {Alfaro}, E.~J.},
        title = "{A non-parametric approach to the membership problem in open clusters.}",
      journal = {\aap},
         year = 1990,
        month = aug,
       volume = {235},
        pages = {94},
       adsurl = {https://ui.adsabs.harvard.edu/abs/1990A&A...235...94C}
}

@ARTICLE{Balaguer2004,
       author = {{Balaguer-N{\'u}{\~n}ez}, L. and {Jordi}, C. and {Galad{\'\i}-Enr{\'\i}quez}, D. and {Zhao}, J.~L.},
        title = "{New membership determination and proper motions of NGC 1817. Parametric and non-parametric approach}",
      journal = {\aap},
         year = 2004,
        month = nov,
       volume = {426},
        pages = {819-826},
          doi = {10.1051/0004-6361:20041332},
archivePrefix = {arXiv},
       eprint = {astro-ph/0407455},
 primaryClass = {astro-ph},
       adsurl = {https://ui.adsabs.harvard.edu/abs/2004A&A...426..819B}
}

@ARTICLE{Javakhishvili2006,
       author = {{Javakhishvili}, G. and {Kukhianidze}, V. and {Todua}, M. and {Inasaridze}, R.},
        title = "{A method of open cluster membership determination}",
      journal = {\aap},
         year = 2006,
        month = mar,
       volume = {447},
       number = {3},
        pages = {915-919},
          doi = {10.1051/0004-6361:20040297},
archivePrefix = {arXiv},
       eprint = {0710.5637},
 primaryClass = {astro-ph},
       adsurl = {https://ui.adsabs.harvard.edu/abs/2006A&A...447..915J}
}

@article{krone2014,
  title={UPMASK: unsupervised photometric membership assignment in stellar clusters},
  author={Krone-Martins, Alberto and Moitinho, Andre},
  journal={\aap},
  volume={561},
  pages={A57},
  year={2014},
  publisher={EDP Sciences}
}

@ARTICLE{Gao2018,
       author = {{Gao}, Xin-Hua},
        title = "{Membership determination of open clusters based on a spectral clustering method}",
      journal = {\pasj},
         year = 2018,
        month = aug,
       volume = {70},
       number = {4},
          eid = {68},
        pages = {68},
          doi = {10.1093/pasj/psy059},
       adsurl = {https://ui.adsabs.harvard.edu/abs/2018PASJ...70...68G}
}

@ARTICLE{Agarwal2021,
       author = {{Agarwal}, Manan and {Rao}, Khushboo K. and {Vaidya}, Kaushar and {Bhattacharya}, Souradeep},
        title = "{ML-MOC: Machine Learning (kNN and GMM) based Membership determination for Open Clusters}",
      journal = {\mnras},
         year = 2021,
        month = apr,
       volume = {502},
       number = {2},
        pages = {2582-2599},
          doi = {10.1093/mnras/stab118},
archivePrefix = {arXiv},
       eprint = {2011.13622},
 primaryClass = {astro-ph.IM},
       adsurl = {https://ui.adsabs.harvard.edu/abs/2021MNRAS.502.2582A}
}

@ARTICLE{deb2022,
       author = {{Deb}, Sukanta and {Baruah}, Amiya and {Kumar}, Subhash},
        title = "{Ensemble-based unsupervised machine learning method for membership determination of open clusters using Mahalanobis distance}",
      journal = {\mnras},
         year = 2022,
        month = oct,
       volume = {515},
       number = {4},
        pages = {4685-4701},
          doi = {10.1093/mnras/stac2116},
       adsurl = {https://ui.adsabs.harvard.edu/abs/2022MNRAS.515.4685D}
}

@ARTICLE{VMG23,
       author = {{van Groeningen}, M.~G.~J. and {Castro-Ginard}, A. and {Brown}, A.~G.~A. and {Casamiquela}, L. and {Jordi}, C.},
        title = "{A machine-learning-based tool for open cluster membership determination in Gaia DR3}",
      journal = {\aap},
         year = 2023,
        month = jul,
       volume = {675},
          eid = {A68},
        pages = {A68},
          doi = {10.1051/0004-6361/202345952},
archivePrefix = {arXiv},
       eprint = {2303.08474},
 primaryClass = {astro-ph.GA},
       adsurl = {https://ui.adsabs.harvard.edu/abs/2023A&A...675A..68V}
}

@ARTICLE{Hayden2015,
       author = {{Hayden}, Michael R. and {Bovy}, Jo and {Holtzman}, Jon A. and {Nidever}, David L. and {Bird}, Jonathan C. and {Weinberg}, David H. and {Andrews}, Brett H. and {Majewski}, Steven R. and {Allende Prieto}, Carlos and {Anders}, Friedrich and {Beers}, Timothy C. and {Bizyaev}, Dmitry and {Chiappini}, Cristina and {Cunha}, Katia and {Frinchaboy}, Peter and {Garc{\'\i}a-Her{\'n}andez}, D.~A. and {Garc{\'\i}a P{\'e}rez}, Ana E. and {Girardi}, L{\'e}o and {Harding}, Paul and {Hearty}, Fred R. and {Johnson}, Jennifer A. and {M{\'e}sz{\'a}ros}, Szabolcs and {Minchev}, Ivan and {O'Connell}, Robert and {Pan}, Kaike and {Robin}, Annie C. and {Schiavon}, Ricardo P. and {Schneider}, Donald P. and {Schultheis}, Mathias and {Shetrone}, Matthew and {Skrutskie}, Michael and {Steinmetz}, Matthias and {Smith}, Verne and {Wilson}, John C. and {Zamora}, Olga and {Zasowski}, Gail},
        title = "{Chemical Cartography with APOGEE: Metallicity Distribution Functions and the Chemical Structure of the Milky Way Disk}",
      journal = {\apj},
         year = 2015,
        month = aug,
       volume = {808},
       number = {2},
          eid = {132},
        pages = {132},
          doi = {10.1088/0004-637X/808/2/132},
archivePrefix = {arXiv},
       eprint = {1503.02110},
 primaryClass = {astro-ph.GA},
       adsurl = {https://ui.adsabs.harvard.edu/abs/2015ApJ...808..132H}
}

@ARTICLE{Qin23,
       author = {{Qin}, Songmei and {Zhong}, Jing and {Tang}, Tong and {Chen}, Li},
        title = "{Hunting for Neighboring Open Clusters with Gaia DR3: 101 New Open Clusters within 500 pc}",
      journal = {\apjs},
         year = 2023,
        month = mar,
       volume = {265},
       number = {1},
          eid = {12},
        pages = {12},
          doi = {10.3847/1538-4365/acadd6},
archivePrefix = {arXiv},
       eprint = {2212.11034},
 primaryClass = {astro-ph.SR},
       adsurl = {https://ui.adsabs.harvard.edu/abs/2023ApJS..265...12Q}
}

@ARTICLE{Zhong22,
       author = {{Zhong}, Jing and {Chen}, Li and {Jiang}, Yueyue and {Qin}, Songmei and {Hou}, Jinliang},
        title = "{New Insights into the Structure of Open Clusters in the Gaia Era}",
      journal = {\aj},
         year = 2022,
        month = aug,
       volume = {164},
       number = {2},
          eid = {54},
        pages = {54},
          doi = {10.3847/1538-3881/ac77fa},
archivePrefix = {arXiv},
       eprint = {2206.04904},
 primaryClass = {astro-ph.GA},
       adsurl = {https://ui.adsabs.harvard.edu/abs/2022AJ....164...54Z}
}

@ARTICLE{Bressan2012,
       author = {{Bressan}, Alessandro and {Marigo}, Paola and {Girardi}, L{\'e}o. and {Salasnich}, Bernardo and {Dal Cero}, Claudia and {Rubele}, Stefano and {Nanni}, Ambra},
        title = "{PARSEC: stellar tracks and isochrones with the PAdova and TRieste Stellar Evolution Code}",
      journal = {\mnras},
         year = 2012,
        month = nov,
       volume = {427},
       number = {1},
        pages = {127-145},
          doi = {10.1111/j.1365-2966.2012.21948.x},
archivePrefix = {arXiv},
       eprint = {1208.4498},
 primaryClass = {astro-ph.SR},
       adsurl = {https://ui.adsabs.harvard.edu/abs/2012MNRAS.427..127B}
}

@ARTICLE{pyupmask,
       author = {{Pera}, M.~S. and {Perren}, G.~I. and {Moitinho}, A. and {Navone}, H.~D. and {Vazquez}, R.~A.},
        title = "{pyUPMASK: an improved unsupervised clustering algorithm}",
      journal = {\aap},
         year = 2021,
        month = jun,
       volume = {650},
          eid = {A109},
        pages = {A109},
          doi = {10.1051/0004-6361/202040252},
archivePrefix = {arXiv},
       eprint = {2101.01660},
 primaryClass = {astro-ph.GA},
       adsurl = {https://ui.adsabs.harvard.edu/abs/2021A&A...650A.109P}
}

@ARTICLE{Cui25,
       author = {{Cui}, Qian and {He}, Zhihong and {Deng}, Shunhong and {Peng}, Liming and {Li}, Chunyan and {Luo}, Yangping and {Wang}, Kun},
        title = "{Census of Blue Straggler Stars in Distant Open Clusters and Maximum Fractional Mass Excess of Open Cluster Blue Straggler Stars}",
      journal = {\aj},
         year = 2025,
        month = apr,
       volume = {169},
       number = {4},
          eid = {219},
        pages = {219},
          doi = {10.3847/1538-3881/adb9e9},
archivePrefix = {arXiv},
       eprint = {2502.18780},
 primaryClass = {astro-ph.SR},
       adsurl = {https://ui.adsabs.harvard.edu/abs/2025AJ....169..219C}
}

@ARTICLE{Skrutskie2006,
       author = {{Skrutskie}, M.~F. and {Cutri}, R.~M. and {Stiening}, R. and {Weinberg}, M.~D. and {Schneider}, S. and {Carpenter}, J.~M. and {Beichman}, C. and {Capps}, R. and {Chester}, T. and {Elias}, J. and {Huchra}, J. and {Liebert}, J. and {Lonsdale}, C. and {Monet}, D.~G. and {Price}, S. and {Seitzer}, P. and {Jarrett}, T. and {Kirkpatrick}, J.~D. and {Gizis}, J.~E. and {Howard}, E. and {Evans}, T. and {Fowler}, J. and {Fullmer}, L. and {Hurt}, R. and {Light}, R. and {Kopan}, E.~L. and {Marsh}, K.~A. and {McCallon}, H.~L. and {Tam}, R. and {Van Dyk}, S. and {Wheelock}, S.},
        title = "{The Two Micron All Sky Survey (2MASS)}",
      journal = {\aj},
         year = 2006,
        month = feb,
       volume = {131},
       number = {2},
        pages = {1163-1183},
          doi = {10.1086/498708},
       adsurl = {https://ui.adsabs.harvard.edu/abs/2006AJ....131.1163S}
}

\begin{appendix}
\section{The extinction coefficient}\label{app:extin_law}
As described in our previous work~\citep{He23b}, for each OC, the extinction coefficient was derived from the polynomial function:

\begin{equation}\label{equ:extinction_coefficient}
\begin{split}
c = c_1+c_2 \ast bp\_rp_0 + c_3 \ast bp\_rp_0^2+c_4 \ast bp\_rp_0^3 \\+ c_5 \ast A_0 +
c_6 \ast A_0^2+c_7  \ast A_0^3 + c_8  \ast bp\_rp_0  \ast A_0 \\+ c_9  \ast A_0  \ast bp\_rp_0^2 + c_{10}  \ast bp\_rp_0  \ast A_0^2
\end{split}
\end{equation}

where $c_1$ to $c_{10}$ values (shown in Table~\ref{tab:extinction coefficient}) were adopted from the public auxiliary data provided by ESA/Gaia/DPAC/CU5 and prepared by Carine Babusiaux.

\begin{table}[ht!]
    \centering
    \caption{The extinction coefficient in different bands}
    \resizebox{\textwidth}{!}{
        \begin{tabular}{ccccccccccc}
            \toprule
             c1 &  c2 &  c3 &  c4 &  c5 &  c6 &  c7 &  c8 &  c9 &  c10 &  Band \\ 
            \hline
            0.66320788 & -0.01798472 & 0.00049377 & -0.00267994 & -0.00651422 & 0.00003302 & 0.00000158 & -0.00007980 & 0.00025568 & 0.00001105 & RP \\ 
            1.15363197 & -0.08140130 & -0.03601302 & 0.01921436 & -0.02239755 & 0.00084056 & -0.00001310 & 0.00660124 & -0.00088225 & -0.00011122 & BP \\ 
            0.99596972 & -0.15972646 & 0.01223807 & 0.00090727 & -0.03771603 & 0.00151347 & -0.00002524 & 0.01145227 & -0.00093691 & -0.00026030 & G \\ 
            0.19404852 & -0.00025957 & 0.00049577 & -0.00026751 & -0.00002923 & 0.00000001 & 0 & 0.00000006 & 0.00000015 & 0 & $K_{\rm s}$ \\ 
        0.34034541 & -0.00150278 & -0.00066457 & 0.00041767 & -0.00015621 & 0.00000018 & 0 & 0.00000342 & -0.00000021 & 0.00000005 & J \\ 
        \bottomrule
        \end{tabular}
    }
    \label{tab:extinction coefficient} 
\end{table}

\section{Astrometric and structural parameters of 45 OCs} \label{app:tab_45ocs}

\begin{table*}
\centering
\caption{Astrometric and structural parameters of the distant OCs analysed in this work.}
\label{tab:this_work}
\setlength{\tabcolsep}{5pt}
\renewcommand{\arraystretch}{1.2}
\resizebox{\textwidth}{!}{
\begin{tabular}{ccccccccccccccc}
\toprule
Idx & Cluster & $l$ & $b$ & $\mu_{\alpha*}$ & $\mu_{\delta}$ & $\varpi$& $\log(\mathrm{Age/yr})$& $A_{\rm v}$& $r_{\rm clu}$& $r_{\rm c}$ & $\kappa$ & $N$ & $N_{\rm G19}$ & Type \\
& & [deg] & [deg] & [mas yr$^{-1}$] & [mas yr$^{-1}$] & [mas] & - & [mag] & [arcmin] & [arcmin] & - &- &- & - \\
(1) & (2) & (3) & (4) & (5) & (6) & (7) & (8) & (9) &(10) & (11) & (12) & (13) & (14) & (15) \\
\midrule
        1 & Arp\_Madore\_2 & 248.123 & -5.885 & -0.50 & 1.27 & 0.09 & 9.70 & 1.68 & 2.62 & 0.43 & 20.01 & 257 & 209 & I \\ 
        2 & Berkeley\_43 & 45.687 & -0.138 & -1.14 & -3.68 & 0.25 & 7.90 & 6.30 & 4.59 & 1.68 & 5.02 & 721 & 399 & I \\ 
        3 & Berkeley\_51 & 72.147 & 0.294 & -3.05 & -4.90 & 0.22 & 8.00 & 5.46 & 2.46 & 0.98 & 6.21 & 529 & 354 & I \\ 
        4 & Berkeley\_52 & 67.897 & -3.174 & -4.25 & -5.67 & 0.18 & 9.25 & 4.46 & 3.97 & 0.95 & 12.57 & 616 & 485 & I \\ 
        5 & BH\_222 & 349.122 & -0.443 & -1.86 & -3.16 & 0.20 & 7.40 & 7.43 & 3.09 & 0.78 & 4.20 & 571 & 279 & I \\ 
        6 & DC\_5 & 286.796 & -0.502 & -5.40 & 2.42 & 0.22 & 8.20 & 3.68 & 3.09 & 1.63 & 3.78 & 634 & 357 & I \\ 
        7 & ESO\_211\_09 & 271.945 & -0.798 & -3.51 & 3.07 & 0.17 & 8.20 & 5.00 & 1.65 & 0.67 & 7.52 & 119 & 77 & I \\ 
        8 & FSR\_0975 & 199.783 & 3.546 & 0.16 & -0.84 & 0.22 & 8.90 & 1.08 & 2.89 & 1.17 & 3.89 & 106 & 36 & I \\ 
        9 & FSR\_1025 & 206.934 & 2.798 & 0.11 & -0.30 & 0.17 & 8.70 & 1.18 & 2.71 & 1.01 & 5.61 & 127 & 66 & I \\ 
        10 & FSR\_1171 & 223.118 & -2.756 & 0.13 & -0.81 & 0.24 & 9.50 & 1.82 & 3.14 & 0.79 & 9.91 & 139 & 60 & I \\ 
        11 & Ivanov\_8 & 254.017 & 0.245 & -2.34 & 3.63 & 0.13 & 8.90 & 3.57 & 1.83 & 0.63 & 11.86 & 163 & 107 & I \\ 
        12 & Juchert\_1 & 47.727 & -0.998 & -1.82 & -4.63 & 0.28 & 8.20 & 5.57 & 2.41 & 1.51 & 3.06 & 252 & 138 & I \\ 
        13 & Kronberger\_79 & 54.178 & -0.613 & -2.74 & -5.36 & 0.14 & 7.00 & 5.70 & 3.62 & 0.80 & 4.06 & 560 & 388 & I \\ 
        14 & Kronberger\_85 & 250.967 & -2.844 & -1.42 & 2.80 & 0.13 & 8.80 & 3.84 & 1.34 & 0.48 & 14.47 & 91 & 56 & I \\ 
        15 & Mayer\_3 & 233.767 & -0.200 & -1.89 & 2.34 & 0.27 & 8.00 & 2.94 & 2.24 & 1.55 & 6.25 & 114 & 66 & I \\ 
        16 & NGC\_1624 & 155.362 & 2.609 & 0.03 & -0.50 & 0.18 & 8.10 & 2.84 & 1.65 & 0.66 & 13.84 & 80 & 41 & I \\ 
        17 & NGC\_3603 & 291.623 & -0.520 & -5.59 & 1.97 & 0.14 & 7.25 & 4.10 & 3.04 & 0.99 & 6.66 & 973 & 469 & I \\ 
        18 & Schuster\_1 & 281.008 & -0.255 & -5.35 & 3.59 & 0.17 & 8.50 & 7.80 & 1.78 & 0.63 & 10.78 & 128 & 95 & I \\ 
        19 & Shorlin\_1 & 290.566 & -0.921 & -5.94 & 2.22 & 0.13 & 7.10 & 4.89 & 1.15 & 0.36 & 3.87 & 140 & 70 & I \\ 
        20 & Westerlund\_1$^\ast$ & 339.547 & -0.402 & -2.22 & -3.73 & 0.28 & 6.50 & 13.50 & 4.01 & 1.30 & 8.17 & 414 & 111 & I \\ 
        21 & Westerlund\_2 & 284.270 & -0.329 & -5.38 & 2.86 & 0.25 & 7.50 & 5.36 & 2.19 & 0.84 & 7.28 & 361 & 159 & I \\ 
        22 & Alessi\_59 & 211.078 & 1.194 & -0.54 & 0.09 & 0.28 & 8.65 & 1.60 & 1.96 & 1.18 & 2.66 & 74 & 17 & II \\ 
        23 & ESO\_313\_03 & 260.427 & -1.307 & -3.08 & 3.71 & 0.19 & 8.60 & 2.54 & 1.95 & 1.51 & 2.54 & 172 & 90 & II \\ 
        24 & FSR\_0524 & 123.623 & -0.744 & -1.60 & -0.31 & 0.27 & 8.40 & 3.28 & 4.10 & 1.85 & 2.48 & 228 & 129 & II \\ 
        25 & FSR\_1580 & 292.147 & 2.636 & -5.28 & 1.36 & 0.15 & 8.80 & 2.44 & 1.16 & 0.55 & 2.46 & 86 & 60 & II \\ 
        26 & Kronberger\_54 & 69.106 & 0.523 & -2.67 & -5.21 & 0.24 & 8.30 & 3.70 & 2.01 & 0.80 & 2.82 & 270 & 163 & II \\ 
        27 & Patchick\_94 & 336.459 & 0.855 & -2.81 & -3.65 & 0.27 & 7.90 & 6.50 & 2.99 & 0.83 & 2.72 & 434 & 289 & II \\ 
        28 & Ruprecht\_32 & 241.574 & -0.574 & -2.03 & 3.14 & 0.20 & 7.60 & 1.30 & 2.60 & 1.34 & 2.49 & 96 & 61 & II \\ 
        29 & SAI\_72 & 213.229 & 1.082 & -0.39 & 0.04 & 0.24 & 8.75 & 1.73 & 2.85 & 2.11 & 2.44 & 103 & 43 & II \\ 
        30 & UBC\_431 & 164.708 & -0.577 & 0.33 & -1.19 & 0.22 & 8.55 & 1.56 & 3.50 & 1.58 & 2.73 & 228 & 83 & II \\ \hline
        31 & CWNU\_542 & 270.241 & -4.964 & -1.98 & 2.43 & 0.11 & 9.10 & 2.41 & 2.01 & 0.41  & 15.13 & 159 & 132 & I \\ 
        32 & CWNU\_543 & 262.758 & -3.235 & -3.13 & 3.53 & 0.14 & 9.45 & 3.08 & 2.82 & 0.52  & 4.42 & 190 & 154 & I \\ 
        33 & CWNU\_544 & 105.565 & 4.185 & -1.65 & -1.53 & 0.15 & 9.35 & 4.25 & 1.73 & 0.81 & 3.29 & 41 & 40 & I \\ 
        34 & CWNU\_545 & 91.332 & 1.492 & -3.15 & -3.68 & 0.25 & 9.15 & 6.48 & 1.89 & 0.57 & 3.47 & 158 & 112 & I \\ 
        35 & CWNU\_547 & 129.282 & 0.765 & -1.57 & 0.61 & 0.17 & 9.15 & 3.84 & 1.82 & 0.66 & 4.87 & 216 & 129 & I \\ 
        36 & CWNU\_548 & 170.445 & 2.281 & 0.08 & -0.46 & 0.25 & 8.10 & 3.01 & 1.98 & 0.77 & 3.86 & 104 & 32 & I \\ 
        37 & CWNU\_549 & 208.454 & 1.653 & 0.27 & 0.36 & 0.20 & 8.75 & 1.72 & 1.33 & 0.34 & 5.21 & 57 & 21 & I \\ 
        38 & CWNU\_550 & 223.219 & -5.089 & -0.37 & 0.89 & 0.12 & 8.95 & 3.10 & 1.46 & 0.32 & 15.43 & 65 & 43 & I \\ 
        39 & CWNU\_553 & 266.915 & -0.368 & -3.15 & 3.68 & 0.16 & 8.70 & 6.30 & 2.11 & 0.72 & 8.12 & 187 & 153 & I \\ 
        40 & CWNU\_556$^\ast$ & 317.123 & 0.166 & -5.94 & -3.58 & 0.21 & 8.00 & 10.00 & 2.71 & 0.84 & 3.13 & 53 & 30 & I \\ 
        41 & CWNU\_546 & 97.076 & 3.609 & -2.72 & -2.45 & 0.14 & 8.20 & 3.00 & 0.90 & 0.48 & 2.20 & 66 & 33 & II \\ 
        42 & CWNU\_551 & 246.125 & 0.357 & -2.00 & 3.11 & 0.22 & 9.05 & 1.65 & 1.83 & 1.07 & 2.17 & 245 & 132 & II \\ 
        43 & CWNU\_552 & 263.061 & -1.168 & -2.93 & 3.46 & 0.20 & 8.00 & 3.50 & 2.67 & 0.89 & 2.72 & 252 & 162 & II \\ 
        44 & CWNU\_554 & 280.213 & -1.578 & -4.82 & 3.61 & 0.18 & 9.20 & 5.90 & 0.84 & 0.58 & 2.95 & 61 & 42 & II \\ 
        45 & CWNU\_555 & 288.799 & -3.370 & -3.86 & 1.94 & 0.17 & 9.35 & 1.23 & 1.03 & 0.64 & 1.95 & 156 & 107 & II \\ 
\bottomrule
\end{tabular}}

\tablefoot{
Columns: (1) cluster index; 
(2) star cluster name; 
(3) and (4) Galactic longitude and latitude; 
(5) and (6) mean proper motions in right ascension and declination; 
(7) mean parallax; 
(8) logarithm of the cluster age in years; 
(9) visual extinction; 
(10) cluster radius; 
(11) core radius; 
(12) density contrast parameter, defined as the ratio of the central stellar density to the background density; 
(13) number of member stars identified in this work; 
(14) number of identified member stars with Gaia $G$-band magnitude $G \geq 19$ mag; 
(15) cluster classification adopted in this work.

$^\ast$ For Westerlund\_1 and CWNU\_556, membership candidates were selected within the core radius ($r \leq r_{\rm c}$); see Section~\ref{sec:field con} for details.}

\end{table*}

\end{appendix}
\end{document}